\documentstyle [epsfig,prc,aps,floats]{revtex}
\begin{document}
\title{ Determination of the pion--nucleon coupling constant and scattering
lengths}

\author{T.E.O.~Ericson\thanks{Also visitor at CSSM, University of
Adelaide, Adelaide 5005, Australia}}
\address{The Svedberg Laboratory, S-75121 Uppsala and CERN,
CH-1211 Geneva 23, Switzerland}
\author{ B.~Loiseau}
\address{Laboratoire de Physique Nucl\'eaire et de Hautes \'Energies 
\thanks{Unit\'e de Recherche des Universit\'es 
Paris 6 et Paris 7, associ\'ee au CNRS},
 LPTPE, Universit\'e P. \& M.
Curie, F-75252 Paris Cedex 05, France} 
\author{A. W.~Thomas}
\address{Department of Physics and Mathematical Physics and 
Special Research Centre for the Subatomic Structure of Matter,
University of
Adelaide, Adelaide 5005, Australia }

\maketitle
 
\begin{abstract} We critically evaluate the isovector GMO sum rule for forward
$\pi$N scattering using the recent precision measurements of $\pi ^-$p and
$\pi^-$d scattering lengths from pionic atoms. We deduce the
charged-pion--nucleon coupling constant, with careful attention to systematic
and statistical  uncertainties.  This  determination gives, directly from data,
$g^2_c({\rm GMO})/4\pi =14.11\pm 0.05$ (statistical)$\pm 0.19$ (systematic) or
$f^2_c/4\pi = 0.0783(11)$. This value is intermediate between that of indirect
methods and the direct determination from backward np differential scattering
cross sections.  We also use the pionic atom data to deduce
 the coherent symmetric and antisymmetric sums of the  pion--proton and
pion--neutron scattering lengths with high precision, namely  $(a_{\pi ^-{\rm
p}}~+~a_{\pi ^-{\rm n}})/2~=~(-12\pm 2$(statistical)$\pm 8$
(systematic))$\times 10^{-4}\ m_{\pi }^{-1}$ and $(a_{\pi ^-{\rm p}}~-~a_{\pi
^-{\rm n}})/2~=~(895\pm 3$(statistical)$\pm 13$ (systematic))$\times 10^{-4}\
m_{\pi }^{-1}$. For the need of the present analysis, we  improve the
 theoretical description of 
the pion-deuteron scattering length. 
\end{abstract}


\section{Introduction}
\label {Introduction}

 The pion--nucleon ($\pi$NN) coupling constant is of fundamental importance in
both nuclear  and particle physics.  In nuclei it sets the scale of the
interaction, together with the pion mass.  In particle physics it is of great
importance for the Goldberger--Treiman  relation   ~\cite {Gol58},  one of the
most important tests of chiral symmetry.  Its experimental error is the main
obstacle in the accurate discussion of the corrections to this relation as
predicted from chiral symmetry breaking  (see, for example, the discussion on
page 1086 of Ref.~\cite{RAH98}).  An accurate test requires a knowledge of the
$\pi $NN coupling constant to a precision of about 1\%, so as to match the
experimental precision of the other quantities in the Goldberger--Treiman
relation. 

The present situation is summarized in Table~I with uncertainties as quoted
by the authors.  In the 1980's, the $\pi $NN coupling constant was believed
to be well known.  In particular, Koch and Pietarinen~\cite{KOC80}
determined a value of the charged-pion coupling constant, 
$g^{2}_c/4\pi =14.28(18)$, from $\pi^{\pm} $p scattering data, 
while Kroll\cite{KRO81}
found the neutral-pion coupling constant $g^2_0/4\pi $ = 14.52(40) from a pp
forward dispersion relation. This was put in question in the early 1990's,
when the Nijmegen group published a series of
papers~\cite{STO93a,STO97,BER90,KLO91} where they reported smaller values on
the basis of energy-dependent partial-wave analyses (PWA) of NN scattering
data. They obtained $g^{2}_0/4\pi =13.47(11)$ and $g^{2}_c/4\pi =13.58(5)$.
Similarly low values with $g^2/4\pi $ about 13.7 have also been found by the
Virginia Tech group~\cite{ARN94a,ARN94b,WOR92,PAV99} from an analysis of
both $\pi^{\pm} $N and NN data.  Using a similar PWA method in the $\pi$p
sector, Timmermans\cite{TIM97} found a value of 13.45(14). These more recent
analyses often suffer from the drawback that they rely on the joint analysis
of large data bases from many experiments with some of the data rejected
according to various criteria.  The statistical accuracy is high, but the
systematic uncertainty is not clear.  Exceptions are the
Goldberger--Miyazawa--Oehme (GMO) sum rule~\cite{GOL55} used by several
groups\cite {ARN94b,SCH99,GIB98} and the forward scattering sum rule for pp
scattering\cite{KRO81}, which, in principle, depend directly on physical
observables.  However, the dominant systematic uncertainties are not
discussed and the uncertainties in the isovector scattering length used as
input are large.  In the case of Ref.~\cite {SCH99} we have corrected
their result as given in Table I to account for an erroneous input value
according to the Erratum of Ref.~\cite {WOR92}.  Another direct
determination is based on the extrapolation of experimental precision data
on single-energy backward differential np cross sections to the pion
pole~\cite {RAH98,ERI95}. This allows a systematic discussion of statistical
and systematic uncertainties, but the uncertainty is so far larger than what
can be achieved at present with the use of the GMO sum rule.  The
extrapolation method gives $ 14.52(26)$, a value significantly larger than
those deduced by indirect methods.  A review of the situation of the $\pi
$NN coupling constant up to 1997 is found in Ref.  \cite{STO97}.  The
problems regarding its determination from np data have recently been
discussed in a dedicated workshop \cite
{Pavan:1999cr,Arndt:1999jc,ERI00,Loiseau:1999fb,Machleidt:1999yd} as well as
in a recent conference working group\cite{SAI99}.

To resolve these discrepancies it is desirable to have an independent precision
determination, directly linked to  measured quantities   
with quantifiable
systematic and statistical errors.  The purpose of the present article is to
demonstrate that recent experimental advances make the GMO relation suitable
for this purpose. The GMO is a forward dispersion relation that expresses the
charged coupling constant $g^2_c/4\pi $ in terms of the isovector $\pi $N
scattering length (70\% contribution) and a weighted integral, $J^-$, of the
difference between the charged-pion total cross sections (30\% contribution). 
This relation has been repeatedly evaluated in the past \cite
{ARN94b,WOR92,SCH99,GIB98,HOE83,ARN95}. Since, until recently, there was little
information on the scattering lengths available from direct data, these
evaluations necessarily relied on scattering lengths extrapolated from
semi-phenomenological $\pi$N phase-shift analyses, using data from a range of
energies above threshold.  At the high precision needed, the systematic errors
in the extrapolated scattering lengths are unclear and have, to our knowledge,
not been estimated. The experimental situation has changed recently.  The
$\pi^-$p and $\pi^-$d scattering lengths can, to high precision, be deduced
from recent experiments on pionic atoms. As a result, all the major ingredients
in the GMO relation can now be discussed as experimentally derived quantities
with transparent sources of uncertainty. Further, the approach can be improved
by the observation that isospin conservation, which was previously assumed, can
be replaced by the weaker assumption of charge symmetry. This avoids the
possibility of perturbations from the rather important violation of isospin
symmetry expected to be associated with the $\pi ^0$p and $\pi ^0$n scattering
lengths \cite {WEI77,FET99}.  The GMO relation can now be completely evaluated
on the basis of data closely linked to direct experiments and it then determines
the charged-pion coupling.   We will develop this
aspect below and also give a discussion of uncertainties in the dispersion
integral.

The paper is organized as follows.  In Section \ref {The GMO sum rule}  we give
a brief review of the GMO sum rule, reorganize it in the most efficient way for
the present purpose, and discuss the magnitudes of the main contributions.
Section \ref {The experimental pi-p and pi-d scattering lengths} presents the
information on the $\pi ^-$p and $\pi ^-$d scattering  lengths deduced from
data on pionic atoms.  We draw the reader's attention to the most critical
 theoretical point in the present procedure for their extraction. Details
on expressions used for the electromagnetic corrections to the 
 experimental $\pi ^-$d scattering lengths are given in Appendix A. 
 In section
\ref {The theoretical pi-d scattering length.}  we  analyze and improve the
theoretical approach to the $\pi $d scattering length  with
 particular attention to a number of smaller terms.
 We use this understanding to deduce
the  most accurate values yet for the $\pi$N scattering lengths  from the
experimental data.  Practical expressions for the  theoretical $\pi ^-$d
scattering length for separable scattering amplitudes are given in 
Appendix B.
In section~\ref {Evaluation of the cross section integral J^- from data.}   we
analyze the uncertainties  from different sources in the cross section integral
$J^-$. In section \ref{Results} we summarize the conclusions about the
scattering lengths and give the GMO sum rule result for the $\pi $NN coupling
constant, $g^2_c/4\pi $, with an  explicit indication of systematic and
statistical uncertainties. 

\section {The GMO sum rule}
\label {The GMO sum rule}

The GMO sum rule for charged-pion--nucleon scattering is a very general forward
dispersion relation, which assumes only analyticity and crossing symmetry. 
Contrary to the usual approach to its evaluation\cite
{ARN94b,WOR92,SCH99,GIB98,HOE83,ARN95}, it is not necessary to assume isospin
symmetry (for a discussion of the GMO relation assuming isospin symmetry see
Eq. (A.6.49) in Ref. \cite {HOE83}). It takes the following form:

\begin {equation}
 f_c^{2}/4\pi = (1-(m_{\pi } /2M)^2)\left [( 1+m_{\pi } /M)
\frac {m_{\pi }}{4}(a_{\pi ^-{\rm p}}-a_{\pi ^+{\rm p}})
- \frac {m_{\pi } ^2}{8\pi^2}\int_{0}^{\infty }
\frac {\sigma^T_{\pi ^-{\rm p}}(k')-\sigma^T_{\pi ^+{\rm p}}(k')}
{\sqrt{k'^2+m_{\pi } ^2}}\ dk'\right ].
\label{eq:GMOrelation}
\end{equation}

Here $m_{\pi }$ is the charged-pion mass and $M$ the proton mass with the
neutron--proton mass difference neglected, $a_{\pi ^{\pm }{\rm p}}$ the 
$\pi^{\pm }$p scattering lengths, $\sigma ^T_{\pi ^{\pm }{\rm p}}$ the total
$\pi ^{\pm }$ proton cross section and $k$ the pion laboratory momentum.  The
relation gives the charged-pion coupling constant  $f_c^2/4\pi
=(m_{\pi}/2M)^2g^2_c/4\pi $  explicitly in terms of the charged-pion scattering
lengths and total cross sections, all directly measurable.  In writing Eq.
(\ref{eq:GMOrelation}) it has been tacitly assumed that Coulomb barrier
corrections have been made to sufficient precision both in the extraction of
the scattering lengths from pionic atoms and, in particular, in the
determination of the total cross sections. We will discuss these issues
as well as   the effect of mass differences and isospin violation further
below.

It is convenient to write the expression (\ref{eq:GMOrelation}) in a simplified
form with numerical coefficients:

\begin {equation}
 g_c^{2}/4\pi  =
-4.50 \times J^- 
+103.3 \times \left (\frac {a_{\pi ^- {\rm p}}-a_{\pi ^+ {\rm p}}}{2}\right ).
\label{eq:GMOsimplified}
\end{equation}
Throughout this paper the scattering lengths are in units of $m_{\pi }^{-1}$,
and $J^-$, given in mb, corresponds to

\begin {equation}
J^-=\frac {1}{4\pi ^2} \int _{0}^{\infty} \frac{\sigma ^T_{\pi
^-{\rm p}}(k')-\sigma
^T_{\pi ^+{\rm p}}(k')}{\sqrt {k'^2+m_{\pi}^2}}\ dk'.
\label{eq:J- definition}
\end {equation}
Everything is in principle measurable to good precision.   The relevant
scattering lengths in Eq. (\ref{eq:GMOrelation}) can be obtained to high
precision using the $\pi ^-$d scattering length as a constraint as will be 
 discussed below. 

So as to obtain a robust evaluation of the coupling constant in the present
context, we rearrange relation (\ref{eq:GMOsimplified}) in such a way that the
most important experimental contributions are explicitly and separately
identifiable: 

\begin {equation}
 g_c^{2}/4\pi  =
-4.50\times J^- +103.3 \times a_{\pi ^- {\rm p}}
-103.3 \times \left (\frac {a_{\pi ^-{\rm p}}+a_{\pi^+{\rm p}}}{2}\right ).
\label{eq:GMOrobust}
\end{equation}

For orientation, and as an initial basis for discussion, we use as a
preliminary value $J^-=-1.077(47)$ mb from Koch~\cite{KOC85} and the
experimental $\pi^-$p scattering length 0.0883(8)$\ m_{\pi
}^{-1}$~\cite{SCH99}.  This gives the following relation, to be improved
later: $g_c^{2}/4\pi = 4.85(22)+9.12(8) -103.3 \times (a_{\pi ^- {\rm
p}}+a_{\pi ^+{\rm p}})/2=13.97(23) -103.3\times (a_{\pi ^- {\rm p}}+a_{\pi
^+ {\rm p}})/2.$ We stress that this is not our final result (our best
estimate of these terms is given in Eqs.~(\ref{eq:a+ determined}),
(\ref{eq:a- determined}) and (\ref{eq:Jminusvalue}) below).  Here the last
term is a small quantity. If we use the old Koch--Pietarinen value~\cite
{KOC80} for $(a_{\pi ^- {\rm p}}+a_{\pi ^+ {\rm p}})/2=a^+= -83(38)\times
10^{-4}\ m_{\pi }^{-1}$ we will find $g_c^2/4\pi=14.83(45)$, while the SM99
solution \cite{ARN99,STR99} with $a^+=20\times 10^{-4}\ m_{\pi }^{-1}$ will
lead to $g_c^2/4\pi=13.76$.  A value for the coupling constant of the order
of 13.6 would require either a relatively large positive magnitude for the
isoscalar scattering length and/or a substantially less negative value for
the cross section integral $J^-$. It is thus extremely important to obtain
an accurate number for the small isoscalar amplitude.  This quantity can be
evaluated with small statistical and systematic uncertainties from the
experimental $\pi ^-$d scattering length, assuming the validity of charge
symmetry, i.e. that the scattering lengths $a_{\pi^+{\rm p}}$ and
$a_{\pi^-{\rm n}}$ are equal. This approximation is expected to be
excellent, since the recent estimate of the isospin violation effect in this
amplitude, mainly due to virtual photon effects \cite {FET99}, suggests that
this leads to an increase of the coupling constant by only 0.2\%. The cross
section integral $J^-$ is at present becoming the largest source of error.
Uncertainties from the small $\pi^-$d term will not have a major impact on
the result. We now turn to a critical discussion of the different
contributions.

\section {The experimental $\pi^-$p and $\pi^-$d scattering lengths} 
\label {The experimental pi-p and pi-d scattering lengths}

The $\pi^-$p scattering length contributes the bulk of the GMO relation and
must be very accurately controlled.  It is deduced from the energy shift in
pionic hydrogen, which (to about 2\%) is proportional to the scattering
length. The highly accurate value from PSI~\cite{SCH99,SIG96} has an
uncertainty dominated by systematics in the analysis.  The accuracy in the
procedure for extracting the scattering length, with a number of small
corrections of electromagnetic origin, has been discussed in detail by Sigg
{\it et al.}~\cite {SIG96a}. The corrections include those for the finite
nucleon and pion size as well as the feedback of the strong interaction
shift on the long-ranged vacuum polarization.  These can all be calculated
to a precision more than an order of magnitude better than the present
experimental error.  They also include the effect of the proton e.m.
polarizability. The crucial step in the analysis is the modeling of the
hadronic interaction. Sigg {\it et al.} have simulated this by using a
short-ranged potential for each of the isospin states with the strength
tuned to the corresponding free scattering length in the absence of the open
$\pi ^0$ channel. They then introduce the open channel via coupled
Klein--Gordon equations and explore the correction for different interaction
ranges, with values near 0.7 $m_ {\pi } ^ {-1}$ that are considered
realistic. The correction and uncertainty are mainly associated with the
conversion between charged and neutral pions due to the available phase
space.  The final theoretical uncertainty is given as 0.5\%, larger than the
statistical uncertainty of 0.2\%.

We have  examined the procedure and agree with  the
quoted electromagnetic corrections and their precision, provided the
hadronic interaction is tuned to correctly reproduce the experimental energy
shift. The treatment of the corrections in the hadronic part, however could
be improved, although it is convincing to a level of a few \%. 

  Lipartia et al.  have demonstrated that Chiral Effective Field Theory
(EFT) gives the same result as the potential approach at least to next to
leading order \cite{LIP01,RUS01} if the physical amplitude is reproduced.
This result is similar to the invariance of the leading order e. m.
correction due to gauge invariance in a energy-dependent potential description\cite
{ERI82}.  It is thus reasonable to simulate the range dependence of the $\pi
$N s-wave amplitude using potentials, provided the low energy expansion of
the s-wave scattering amplitude $f_{0}$ is correctly reproduced to order
q$^2$.  This latter approach automatically includes the wave function
modification by the extended charge distribution, an effect of higher order
in the EFT approach, but which gives here the largest numerical correction.
However, the procedure in Ref.~\cite {SIG96a} does not respect the empirical
values for the `range' terms, which leads to a larger uncertainty than the
one quoted for their correction.  The negative sign of the correction term
and its approximate magnitude of $-1 $\% is basically correct. To account
for the present inconsistency with the range expansion and using the
numerical range of variation of Sigg et al., the theoretical
uncertainty must be increased from $\pm 0.5\%$ to $\pm 1.0\%$, i.e.  the
overall systematic uncertainty in the scattering length  taken in
quadrature  is increased from $\pm  6\times 10^{-4} m_{\pi
}^{-1}$ to $\pm 10\times 10^{-4} m_{\pi }^{-1}$. We have not
attempted to correct the deduced scattering length of Ref.  \cite{CHA97},
since this should be investigated specifically
\cite {GAS00,ERI00a}.  Range corrections to  the $ \pi ^-$p  width are
not relevant at present accuracy.

The isospin breaking in the $\pi ^-$p amplitude has been dimensionally
estimated in chiral EFT theory\cite {RUS01}.  Such effects are modeled in
the potential approach as well.  The estimate in EFT in next to leading
order appears to be a considerable overestimate owing partly to higher order
compensations. The main uncertainty in the estimate of Ref.
\cite {RUS01} is absent in the difference between the $\pi ^{\pm }$p
amplitudes, which is  the quantity relevant to the GMO relation for
the $\pi $NN coupling constant.

The experimental $\pi^-$d scattering length is derived from the energy shift
in the $\pi^-$d atom in close analogy to the case of the $\pi^-$p scattering
length. The deuteron electromagnetic corrections can in practice be
calculated using a deuteron charge distribution, that correctly reproduces
the experimental deuteron charge radius.  Further, the deuteron is simpler
in so far as the the correction for the open $\pi ^0$ channel is negligible.
The electromagnetic corrections produced to the the strong $\pi $d amplitude
should be included, however.  The main one originates in the energy
dependence, similar to the case of the proton.  This small, repulsive
contribution to the energy shift can be estimated to leading order from our
approach in Ref.\cite{ERI82}, Eqs. (3-5) and it is mainly produced by the
leading order isoscalar range term (see Appendix A). The estimated change in
the deduced scattering length is $-4m_{\pi }b^+ e \langle V_C^d(r) \rangle
$, where the Coulomb potential from the extended deuteron charge
distribution is averaged over the deuteron matter distribution.  Note that
there are no cancellations in the range terms, contrary to the massive
cancellation of the $\pi $N scattering lengths in the single scattering term.
Numerically, the empirical value for the range terms are
$b^+=-0.044(7)m_{\pi }^{-3}; b^-=0.013(6)m_{\pi }^{-3})$\cite {HOE83}.  Any
modern deuteron density distribution gives $e \langle V^d_C(r) \rangle
=0.86$ MeV and a correction of $12\times 10^{-4} m_{\pi }^{-1}$.  An
alternative estimate is obtained from the gauge correction to the $\pi ^-$n
amplitude due to the Coulomb field of the proton, treated as a static
spectator. Using the empirical $\pi $N range parameters this gives a
contribution $-2m_{\pi }(b^+-b^-) e \langle V^p_C(r) \rangle = 6\times
10^{-4} m_{\pi }^{-1}$ with $e\langle V^p_C(r) \rangle =0.66$ MeV.  A
related estimate in a leading order chiral approach gives a correction
$7.5\times 10^{-4} m_{\pi}^{-1}$\cite {ROC95}, but it is based only on the
isovector term and does not include the constraints of the phenomenological
range expansion.  In the absence of correlations between the nucleons, the
isovector range term does not contribute to leading order and it is further
suppressed by its empirical weakness.  We adopt the average of the first two
estimates of $9\times 10^{-4}m_{\pi }^{-1}$ for this correction with an
uncertainty of $5\times 10^{-4}m_{\pi }^{-1}$.  This is well inside the
present uncertainty in the theoretical deuteron scattering length (see Table
IV) and has little influence on the present investigation. 

In summary, we have adopted the following  scattering lengths deduced from
 the data on the $\pi ^-$p atom\cite {SCH99,SIG96a} and the $\pi ^-$d 
 atom\cite {HAU98}
 with the modifications
described above.   The
transition amplitude $a_{\pi ^-p\rightarrow \pi ^0 {\rm n}}$  is the one obtained from the
width of the 1s state of the $\pi ^-$p atom\cite {SCH99,SIG96a}:

\begin {equation} 
a_{\pi ^-p\rightarrow \pi ^-{\rm p}}=~ (883\pm 2{\rm (statistical)}\pm 10
{\rm (systematic)})\times 10^{-4}
\ m_{\pi }^{-1},
\label{eq:exp pi-p scattering length}
\end {equation}
\begin {equation} 
a_{\pi ^-p\rightarrow \pi ^0 {\rm n}}=
~1280(60)\times 10^{-4} \ m_{\pi }^{-1}, 
\label{eq:exp pi-p pi0n scattering length}
 \end {equation}
\begin {equation} 
a_{\pi ^-{\rm d}}~=~(-252 \pm 5 {\rm (statistical)} \pm 5 
{\rm (systematic)}+i63(7))\times 
10^{-4}\ m_{\pi }^{-1}.
\label{eq:exp pi-d scattering length} 
\end {equation}
We recall that  the following relations hold, if isospin symmetry is assumed to
be valid: $a_{\pi ^-p}\equiv a_{\pi ^-p\rightarrow \pi ^-{\rm p}}=a^+~+ ~a^-~~;
$ $~~  a_{\pi^-p\rightarrow \pi ^0 {\rm n}}=-\sqrt{2}a^-$, 
 where $a^{\pm }$ are the symmetric and antisymmetric scattering lengths 
$a^{\pm }=\frac {1}{2}(a_{\pi ^- {\rm p}}\pm a_{\pi ^+ {\rm p}})$, 
respectively.

\section {The theoretical $\pi^-$d scattering length}
\label {The theoretical pi-d scattering length.}

The part of the GMO relation, Eq. (4), that it has not been possible to
determine accurately up to now is the term proportional to the coherent,
symmetric combination of the scattering lengths  $(a_{\pi ^-{\rm p}}+a_{\pi
^+{\rm p}})/2$. Assuming isospin symmetry, this is the isoscalar scattering
length $a^+$. It follows from recent measurements of the hadronic energy shift
and width of the pionic hydrogen  atom~\cite {SCH99} that this gives a directly
determined value  $a^+=-22(43)\times 10^{-4}\ m_{\pi }^{-1}$. However, the
accuracy of this direct determination is not sufficient for our present
purpose. It is very difficult to determine $a^+$ with precision, directly from
the coherent sum of the individual $\pi ^-$p and $\pi ^+$p scattering lengths,
because these cancel to a few per cent . On the other hand, assuming only
charge symmetry, this quantity is identical to the coherent scattering length
for a negative pion on the neutron and proton,  $(a_{\pi ^-{\rm
p}}+a_{\pi^-{\rm n}})$/2,  which is the leading contribution to the accurately
known $a_{\pi ^-{\rm d}}$ scattering length. The accuracy of this approximation
is indicated by a recent estimate of the isospin violation effect in the
amplitude ratio $R_4=-0.008(1)$~\cite {FET99} such that

\begin {equation} a_{\pi ^+{\rm p}}
-a_{\pi ^-{\rm n}}=R_4\ a_{\pi
^-{\rm n}}=3\times 10^{-4}\ m_{\pi }^{-1}.
\label{eq:pi-p pi+p charge symmetry violation}
\end {equation}

Provided the remaining contributions can be reliably calculated, it is then
possible to deduce the relevant coherent combination directly from the deuteron
data with only minor assumptions concerning isospin symmetry.  The situation is
exceptionally favorable for the application of multiple scattering methods. The
deuteron is a very loosely bound system and its wave function is accurately
known.  The nucleons have very little overlap and, consequently, the poorly
controlled short range contribution is small.  The particular case of the
$\pi$d scattering length is even a textbook example of multiple scattering (see
p. 111 in Ref.\cite{ERI88}),  since the expansion parameters are small.  The
situation has been explored in detail, both within multiple scattering theory
and using a three-body Faddeev approach, since it provides a clear-cut testing
ground for methods\cite{AFN74,MIZ77,THO80,FAL77,BAR97a}.

In the static (fixed scattering centers) approximation the leading structure
and scale of the pion--deuteron scattering length is set by the coherent single
scattering term $S$ and the dominant s-wave double scattering term $D$  which
is proportional to the inverse deuteron radius $\langle 1/r \rangle $  (p. 111
in Ref. \cite {ERI88}): 

\begin {equation} 
a_{\pi ^-{\rm d}}^{{\rm static}}~= ~ S~+~D~ ...;
\label{eq:pi-d scattering length schematic}
\end {equation}
\begin {equation}
S~=~\frac {(1+m_{\pi}/M)}{(1+m_{\pi}/M_{{\rm d}})}
\ (a_{\pi ^-{\rm p}}+ a_{\pi^-{\rm n}});
\label{eq:pi-d single scattering}
\end {equation}
\begin {equation}
D~=~2~\frac {(1+m_{\pi}/M)^2}{(1+m_{\pi}/M_{{\rm d}})}~
 \left [ \left (\frac {a_{\pi ^-{\rm p}}+
a_{\pi^-{\rm n}}}{2}\right )^2-2 \left (\frac {a_{\pi ^-{\rm p}}-
a_{\pi^-{\rm n}}}{2}\right )^2 \right ]\langle 1/r\rangle ,
\label{eq:pi-d double scattering} 
\end {equation}
where $M_{{\rm d}}$ is the deuteron mass.

The static double scattering term  represents about 90\% of the experimental
scattering length. It is in practice well defined numerically with a small
error from the uncertainty in $<1/r>$. It has
 typically the value 

\begin{equation} D=-254(4)\times 10^{-4}\ m_{\pi}^{-1},
\label{eq:pi-d double scattering numerical}
\end {equation}
where we have used the central values of the scattering lengths from Eqs. (20--21).
 We will use this well defined static limit with point
interactions as the starting point with respect to which various corrections
will be introduced.

\subsection { Previous approaches to $a^+$ from the deuteron data}  

Recently Baru and Kudryatsev (B--K)~\cite{BAR97a} have investigated the
$\pi$d scattering length using state-of-the-art multiple scattering methods.
We will use the updated and unpublished version of their investigation
\cite{BAR98} as the theoretical yardstick for the following discussion.  We
have numerically reproduced their findings to the same numerical precision,
under the same assumptions.  This approach is however still incomplete and
contains, we believe, one erroneous term. As a consequence, the close
agreement of their quoted value
$a^+=-15(9)\times 10^{-4}\ m_{\pi}^{-1}$ with our final result for a$^+$
is only a fortuitous numerical coincidence without any special significance.
It cannot be used as such.
 In the following we discuss 
the input parameters, corrections and systematics, and introduce substantial
theoretical improvements. The classical 3-body approach to the problem is
still that of Afnan and Thomas and of Mizutani and Koltun, using separable
interactions~\cite {AFN74,MIZ77}. This approach gives the best picture of
the dispersive effects due to absorption and supports the conclusions of the
heavy cancellation of unitarity corrections in the multiple scattering
approach.  The approach, however, has not been updated in its overall
accuracy to match the present high experimental precision and cannot be used
directly.

A rather different approach is that of Beane {\it et al.}~\cite {BEA98}, based
on the nuclear chiral perturbation approach of Weinberg~\cite{WEI92} and using
phenomenological deuteron wave functions. This approach makes a systematic
expansion in the pion 4-momentum, using effective parameters; at present the
calculations have been made to O(q$^3$). The result has the same general
structure as the static limit of multiple scattering.  
 Several  physical effects  discussed  in the  following  are not  yet
included  in  this  order, such  as  the  Fermi  motion term  and  the
dispersive  correction  from  pion  absorption.   They  conclude  that
$a^+=-30(5)\times  10^{-4}\  m_{\pi}^{-1}$   to  O(q$^3$),  where  the
uncertainty  represents  only  the  experimental  uncertainty  in  the
deuteron scattering length. The systematic uncertainty from the omitted
higher  order terms  is  most  likely nearly  one  order of  magnitude
larger.

 \subsection { The inverse deuteron radius} 

The inverse deuteron radius appearing in Eq. (\ref{eq:pi-d double scattering})
must be evaluated from wave functions. It is essential that the asymptotic
normalization be accurately consistent with the experimental np effective range
and that the wave functions correspond to an energy-independent interaction.
The Paris \cite {LAC81} and Bonn2 \cite {MAC87} wave functions satisfy these
criteria and give $\langle 1/r\rangle _{{\rm Paris}}$ = 0.449~fm$^{-1}$ and
$\langle 1/r\rangle _{{\rm Bonn2}}$ = 0.463~fm$^{-1}$ with asymptotic
normalizations $A_S$(Paris) = 0.8869~fm$^{-1/2}$ and $A_S$(Bonn2) =
0.8863~fm$^{-1/2}$, respectively.  We have conservatively used the average of
these model values $\langle 1/r\rangle $ = 0.456(7) fm$^{-1}$~=~0.645(10)$\
m_{\pi }$; the uncertainty given is set by their difference. We note that the
inverse radius, 0.520 fm$^{-1}$, of the Hulth\'en wave function\cite {FAL77},
which is often used for explorations of various effects, is nearly 15\% larger
than these values and should not be used in quantitative studies. The
uncertainty in the theoretical $\pi $d scattering length from the inverse
radius is less than its present experimental precision.

\subsection { Effects of the non-locality of the $\pi $N s-wave
interaction} 

The simplest approximation to the double scattering  term of Eq. (\ref{eq:pi-d
double scattering}) assumes that the $\pi $N scattering is point-like. Such an
approximation is appropriate if the two scatterers are well separated, as is
the case for the bulk of the contributions in the case of the deuteron as a
consequence of its loose binding.  The rather small non-local correction must,
however, be controlled in sign and magnitude at the level of precision aimed
for here.  However, it is not necessary to describe this effect very
accurately. The non-local effects enter mainly in the description of the
isovector $\pi $N s-wave interaction, which is well known to be closely
associated with $\rho $-meson exchange and which heavily dominates the double
scattering term. For calculational convenience it has been conventional to
model the non-locality of the scattering amplitude in terms of a separable
form, $v(k)v(k')$, with a monopole form factor  $v(k)=c^2/(c^2+k^2)$. Since the
initial and final pion are at rest with momentum 0 and the intermediate pion
has momentum {\bf q}, this means that in momentum space the static pion
propagator changes from $q^{-2}$ to $v(q)^2q^{-2}$.  In coordinate space this
corresponds to a change of the expectation value $\langle 1/r\rangle $ by

\begin{equation}\delta
\langle 1/r\rangle =
- \left \langle \left |\frac {1+c\ r/2}{r}\exp(-cr)\right |\right \rangle  
\label{eq:deltaoneoverr}.
\end{equation}
We list in Table II the values of $\delta\langle 1/r\rangle $ and the
corresponding contribution to the deuteron scattering length for different
values of $c$ as well as the contribution to the scattering length for standard
values of the $\pi $N scattering lengths. 

 We conservatively consider  that plausible values for the
parameter $c$ lie in the interval  $3.5\le c \le 5\ m_{\pi }$. This is
a wide  range, which should  adequately cover any model  dependence of
the  result. These  values have  been obtained  using two  extremes of
strong form factors  for the double scattering term.  One choice is to
consider each of the scatterings to be associated with a monopole form
factor. Since  the isovector scattering strongly  dominates the double
scattering,  the   natural  cut-off  parameter   is  the  $\rho$-meson
mass.  This would  give the  same correction as  quoted in Table  II for
$c=5\ m_{\pi }$.  Another choice  is include in addition a strong form
factor  of typical  $\rho  $-meson range  for  both the  pion and  the
nucleon.   The effective  overall  form  factor in  each  of the  pion
scatterings is then a dipole  form factor with the $\rho $-meson mass,
corresponding to $c=3.5\ m_{\pi }$.
 It should be observed that the typical modification of $\langle 1/r\rangle
$ is a negative contribution by 4 to 8\% corresponding to a positive
contribution to $\delta a_{\pi ^-{\rm d}}$ of 9 to 20  $\times 10^{-4} \
m_{\pi}^{-1}$.  We choose the mean of these two approaches as a typical value
with the spread setting the scale of the uncertainty, but note that in doing so
we may somewhat underestimate the non-local effect, such that our final value 
of $g_c^2/4\pi $ may be somewhat  too low.

 We found that the results reported by B--K in Ref. \cite
{BAR97a}, Table 3, for the realistic Bonn1 and Bonn2 wave functions did not include
the form factor (contrary to the statement in the paper), which the authors
 confirm.  We have received their corrected and extended results \cite {BAR98}
 for the Bonn1 potential.  
 Note that at the present level of
precision it is important to use potentials fully consistently.  The Bonn1
potential is energy-dependent; as a consequence, orthonormality can only be
respected in matrix elements calculated using this potential if non-trivial
weight factors are introduced in the integrands.  To eliminate this
uncertainty we use here the similar, but energy-independent, Bonn2
potential.  B--K consider without arguments cut-off values $c=2.5, ~3$ and
$3.5 \ m_{\pi}$ in the form factor; this gives positive contributions to the
scattering length as compared to the point-like static approximation of 36,
27 and 22 $\times 10^{-4} \ m_{\pi}^{-1}$ , respectively.  
  There are
 good physical reasons to believe that $\rho$-meson 
exchange sets
the scale for the dominant isovector amplitude with a larger value for the
effective $c$. To be conservative we take $c= 3.5$ and $c=5 \ m_{\pi}$ for
the cut off as limits for this systematic correction from the non-localities
and use the central value of these two extremes as the correction. This is
smaller than the correction. Our correction is smaller  than the one 
 found by B--K. Non-locality is one of the
largest theoretical sources of systematic uncertainty in corrections to the
point-like static approximation.

\subsection {Corrections to the static approximation} 

The nature of the leading non-static corrections and the reasons why the static
expression (fixed scattering centers) still remains an excellent approximation
are well understood. At first sight, even the single scattering amplitudes have
rather important non-static modifications, representing about 30\% of the total
$\pi $d scattering length. Such corrections are systematically generated by the
multiple scattering description in which physical amplitudes are used, thus
guaranteeing the correct behavior of the scattered wave at large distances. The
emphasis is thus not on the near-zone behavior of the scattering as in
pseudo-potential or effective Lagrangian approaches. In a situation like the
present one, this leads to a systematic cancellation of unitary binding
corrections between single scattering and double scattering terms, when these
are introduced consistently. This phenomenon was first demonstrated in the
present context for an analytically soluble model by F\"aldt \cite {FAL77}.  It
has been numerically investigated by B--K \cite {BAR97a} using a Hulth\'en wave
function and a separable amplitude with a dipole form factor and a cut-off
parameter 3$\ m_{\pi}$. They conclude that the amplitude increases by only $10 
\times 10^{-4}\ m_{\pi}^{-1}$, when the non-static term is included.  This is
only twice the experimental uncertainty and less than the uncertainty from the
form factor. F\"aldt evaluated the joint contribution of the non-static and the
form factor terms using a dipole form factor with c=3.6$\ m_{\pi }$ with a
Hulth\'en wave function \cite {FAL77}. The overall contribution corresponds to
34$\times 10^{-4} \ m_{\pi }^{-1}$. The comparison with our independent
evaluation of pure form factor corrections indicates that the non-static term
in this case is about 8$\times 10^{-4}\ m_{\pi }^{-1}$. A detailed calculation
of this correction is complicated. Wycech informed us that he is in the process
of reevaluating the non-static contributions using a Faddeev approach and
separable interactions. At the present moment he has only results using an
interaction that reproduces the Hulth\'en wave function; this gives +12$\times
10^{-4}\ m_{\pi }^{-1}$, in excellent agreement with the previous
results~\cite{WYC00}.  Following B--K we have adopted a value 11(6)$\times
10^{-4}\ m_{\pi}^{-1}$, where the liberal uncertainty reflects the lack of
verification of the value of non-static effects using high quality deuteron
wave functions.

\subsection { Fermi motion}

Another well defined correction originates in the nucleon Fermi motion.  In the
case of s-wave scattering, such contributions cancel systematically to high
precision with other binding terms \cite {FAL77}.  In addition, the single
scattering term from the $\pi$N p-wave scattering produces a small, attractive
and  physically well understood contribution, which can be reliably evaluated
as a leading order effect originating in the nucleon momentum distribution and
the spin--isospin averaged p-wave threshold scattering amplitude
  $c_0$=0.208$\ m_{\pi}^{-3}(3)$\cite {ERI88}. 

\begin{equation}  
a({\rm Fermi})~=~ 2\ c_0
\frac {m_{\pi }^2(1+m_{\pi}/M)}{(M+m_{\pi})^2(1+m_{\pi}/M_{\rm d})}
\ \left\langle p^2\ v^2\left(\frac {m_{\pi }}{M+m_{\pi }}p\right)
\right\rangle.
\label{eq:Fermimotion}
\end{equation}

We  have  calculated  this  expectation  value for  two  high  quality
deuteron wave functions. The results are given in Table III.  The form
factors  are  manifestly  of   no  importance.  The  relatively  large
difference between the Paris  potential and the Bonn2 potential arises
because  of the  D-state component,  which generates  contributions 10
times more  effectively than  the S-state one.  The difference  in the
correction in the  two cases is thus almost  entirely a consequence of
the  well  known  difference  in  the  D-state  probability  (P$_{{\rm
D}}$=5.7\%  vs. 4.3\%)  for the  two wave  functions.   The normalized
momentum distributions for the  S- and D-wave component, respectively,
are very similar in the two  models.  We therefore treat its effect as
a true model dependence. We take the spread in the values of the Fermi
motion  corrections   as  a   measure  of  a   systematic  theoretical
uncertainty,  although  physical arguments  for  the  higher value  of
P$_{{\rm  D}}$  exist \cite{ERI85}.   Consequently,  in the  following
evaluation,  we  use the  value  $a({\rm Fermi})=61(7)\times  10^{-4}\
m_{\pi}^{-1}$. This is larger than the value 50 to $53 \times 10^{-4}\
m_{\pi }^{-1}$ found by B--K based on the Bonn1 and 2 wave functions.

\subsection { Dispersion contribution} 

A small repulsive contribution, not described by multiple scattering, is
produced by the dispersive term from the absorption reaction $\pi ^-$d
$\rightarrow $ nn. This quantity has been repeatedly calculated using Faddeev
approaches \cite {AFN74,MIZ77,THO80}. It typically has a theoretical
uncertainty of 20\% of its numerical value $-56(14) \times 10^{-4}\
m_{\pi}^{-1}$ \cite{THO80}.  The dispersive contribution is a theoretically
calculated correction; a more detailed study of this term is highly desirable.
The uncertainties reflect the model dependence of the approach.

\subsection { sp interference}

This is the name given by B--K to a term originating in pion p-wave scattering
on one of the nucleons due to Galilean invariance \cite{BAR97a}. Such Galilean
terms generate s-wave scattering contributions even for pion scattering on free
nucleons. In the present situation the relevant spin-averaged on-shell
scattering volume for charge exchange of a p-wave pion is well known and the
corresponding scattering amplitude on-the-mass-shell depends on the pion
momentum in a well defined way.  The Galilean correction for nucleon motion
involves going off the mass shell and usually depends on the description. B--K
advocate that a contribution of about $ 42 \times 10^{-4}\ m_{\pi }^{-1}$
originates from p-wave scattering due to the momentum of the intermediate pion
when expressed in the $\pi $N CM system.  However, in the present situation the
contribution is almost entirely generated by the isovector $\pi$N Born term and
it can be evaluated exactly. From the expressions given in H\"ohler's reference
book, Eq. (A.8.2)~\cite {HOE83}, one finds that this term is proportional to

\begin{equation}
\nu ^2-\frac {(k^2+k'^2-t)}{2}=\nu ^2-q\cdot q'.
\label{eq:spinterference}
\end{equation}
Here, $\nu$ is (to order $M^{-2}$) the Breit frame pion energy, which is
proportional to the scalar product of the average 4-vectors of the nucleons
($p$ and $p'$) and pions ($q$ and $q'$), respectively.

\begin{equation}
\nu =\frac{1}{M}\frac {(p+p')}{2}\cdot \frac {(q+q')}{2}=
\frac{(q_0+q_0')}{2}-\frac {1}{M}
\frac {({\bf p+p'})}{2}\cdot \frac {({\bf q+q'})}{2}.
\label{definition nu}
\end{equation}
(Eq. (A.1.6) in  Ref.~\cite {HOE83}). Thus, neglecting
terms of order $M^{-2}$, the pion pole term is proportional to
\begin{equation}
\frac {(q_0-q_0')^2}{4}-\frac {(q_0+q_0')}{M}\frac {({\bf
p+p'})}{2}\cdot
\frac {({\bf q+q'})}{2}+{\bf q\cdot q'}.
\label{eq:sp interference pion pole}
\end{equation}

In the double scattering term, the contribution comes from nucleon 1 with
initial (final) momentum ${\bf p} ~({\bf p-q'})$ and with the initial
(intermediate) pion momentum ${\bf 0} ~({\bf q'})$, respectively, while for
nucleon 2 the initial (final) nucleon momentum is ${\bf -p}~({\bf -p+q'})$ with
intermediate (final) pion momentum ${\bf q'}~({\bf 0})$, respectively; the pion
energies, $q_0$ and $q_0'$, are unchanged in this term.  The sum of these two
contributions are

\begin{equation}
\frac {q_0}{M}\frac {{\bf q'}^2}{2}-\frac {q_0}{M}\frac
{{\bf q'}^2}{2}=0.
\label{eq:sp double scattering pion term}
\end{equation}

On the other hand, B--K  make the choice of Galilean invariance {\it for the
incoming and outgoing $\pi $N systems calculated  separately} in the primary
amplitude and find in the same limit $ 0+q_0{{\bf q'}^2}/M$ in Eq. (\ref{eq:sp
double scattering pion term}). Instead the  exact pole term corresponds, to
order $M^{-2}$, to a Galilean invariant expression using the {\it average}
velocity of the initial and final nucleons, contrary to the B--K assumption. In
other words, the pole term is proportional to the scalar product of the pion
momenta ${\bf q}_{B}\cdot {\bf q'}_{B}$ in the nucleon Breit frame. We have
therefore suppressed this term in the  B--K  multiple scattering expansion.

We note in passing that, even if the Galilean contributions were of the type
proposed by B--K, their importance would most likely be strongly suppressed.
The reason is that these terms generate a $\delta $-function interaction in the
absence of form factors.  We therefore suspect that NN correlations would
largely suppress such contributions, in analogy with the
Ericson--Ericson--Lorenz--Lorentz effect for p-wave $\pi $ propagation in the
nuclear medium (p. 140ff in Ref.~\cite {ERI88}).

\subsection { Isospin and mass difference corrections} 

In the above expressions, we assumed that isospin holds for the calculation
of double scattering and that charge symmetry holds for the single
scattering.  We now quantify the effect of these approximations. B--K have
investigated the consequence of the physical mass difference between $\pi
^-$ and $\pi ^0$ and between the neutron and the proton in the multiple
scattering.  They find an increase of the scattering length by about
$3.5\times 10^{-4}\ m_{\pi }$. The smallness of this term is in part due to a
systematic compensation of single and double scattering contributions in
analogy to the compensation of unitarity corrections to single and double
scattering terms.  As an alternative approach we use the recent estimates of
the violation of isospin symmetry from light quark mass differences and
virtual photon effects in the $\pi $N scattering lengths
\cite{FET99}.  We maintain only the effects of violations in the
amplitudes in the double scattering term in view of the systematic
cancellation between single scattering and propagator modifications in the
double scattering term.  This leads to an increase of the scattering
amplitude by $3.5\times 10^{-4}\ m_{\pi }$, numerically identical to the
previous estimate. It is not clear whether these approaches represent the
same physics and this point should be further investigated. However, both
results indicate that the effects are small in the present context,
although they will become of importance in the future.  In view of its
smallness and since it is not at present established experimentally,
we have not included this correction, which is within experimental
uncertainties. It has, however, been included as an uncertainty in our
estimate of systematic errors.

 \subsection { Higher order multiple scattering corrections} 

In the present case the multiple scattering expansion is rapidly convergent
beyond the double scattering term. In the fixed scattering approximation with
separable interactions, these higher order terms can be summed exactly to all
orders.  B--K calculated these terms approximately, assuming point-like
scatterers. We have verified these calculations and reproduce their results.
They have since improved the evaluation of this small term, using form factors
and find a stable contribution to the scattering length of the order of
$+6\times 10^{-4}\ m_{\pi}^{-1}$ \cite {BAR97a,BAR98}. Our independent
evaluation also gives very stable values, but somewhat smaller, in the range of
3 to $4\times 10^{-4}\ m_{\pi}^{-1}$ for the form factors considered.  We have
used the value $4(1)\times 10^{-4}\ m_{\pi}^{-1}$ for this correction. The
effect is much smaller than other uncertainties, for example those due to form
factors.

\subsection {Inverse pion photo-production} 

Another small electromagnetic correction comes from the physical s-wave
photo-production process $\pi ^-{\rm p}\rightarrow \gamma $n  acting on one
nucleon followed by the inverse reaction on the other one. This double
scattering process has nearly the same structure as the corresponding s-wave
charge exchange process  $\pi ^-{\rm p}\rightarrow \pi ^0$n  in Eq.
(\ref{eq:pi-d double scattering}), but for the fact that the intermediate
photon now has momentum  k$_{\gamma }=m_{\pi}$ in the static limit, such that

\begin{equation}{\rm Re}
\ D_{\gamma }=-2/3\ \frac {(1+m_{\pi }/M)^2}{\ (1+m_{\pi }/M_{{\rm d}})} 
\ \left[E_{0+}(\gamma {\rm n}\rightarrow\pi ^-{\rm p})\right]^2
\ \left\langle \frac {\cos(k_{\gamma }r)}{r}\right\rangle .  
\label{eq:photoproduction Dgamma}
\end{equation} 
 Here the photo-production amplitude  $E_{0+}(\gamma {\rm n}\rightarrow \pi
^-{\rm p})=-31.4\times 10^{-3}\ m_{\pi }$ (Table 8.3 in Ref.~\cite{ERI88}). 
 This small charge dependent term is of order $-2 \times
10^{-4}\ m_{\pi }^{-1}$, which  is a magnitude less
 than the overall theoretical uncertainty; 
cf. also Ref. \cite{ROC95}.

\subsection {Double p-wave scattering} 

A small correction results from the p-wave scattering due to nucleon motion at
both vertices.  This effect has been estimated by B--K for an analytically
soluble deuteron model with Gaussian wave functions.  They find a contribution
of    about $-3\times 10^{-4} \ m_{\pi}^{-1}$.  We have included this small
effect.

\subsection { Scattering on virtual pions} 

Finally, one may envisage a contribution from the scattering of the pion on a
virtually exchanged pion in the deuteron.  However, we are dealing with an
isoscalar system, and such a contribution is proportional to virtual isoscalar
$\pi \pi $ s-wave scattering and should be very small, from a chiral
perspective. In particular, since the deuteron is such a loosely bound system,
one expects this term to be small. Robilotta and Wilkin showed that large
cancellations in a consistent treatment give only  $-5\times10^{-4}\ m_{\pi
}^{-1}$~\cite {ROB78}.  This is confirmed by a recent chiral estimate of $-8$
to $-6\times 10^{-4}\ m_{\pi }^{-1}$~\cite {BEA98}. We adopt a contribution of
$(-6 \pm 2)\times 10^{-4}\ m_{\pi }^{-1}$ from this effect.

\subsection {Results for the $\pi $N scattering lengths}

The different contributions from the previous subsections  are summarized  in 
Table IV, using the final parameters from Eqs.(\ref{eq:a+ determined}) and
(\ref{eq:a- determined}) whenever  appropriate. Consequently, the present 
energy shift in the $\pi ^-$d atom leads to the following value for the 
coherent scattering length from a proton and a neutron:

\begin {equation}
\frac {a_{\pi ^-{\rm p}}+a_{\pi ^-{\rm n}}}{2}=
(-12\pm 2  {\rm (statistical)}\pm 8{\rm (systematic)})\times
10^{-4}\ m_{\pi}^{-1}.
\label{eq:a+ determined}
\end {equation}

The high accuracy  is a direct consequence of the the very strong constraint provided by the  $\pi ^-$d atom level shift. The usual determination via phase shift analysis is difficult, since it requires  differences between large numbers.
In the limit of isospin symmetry, this quantity is the isoscalar scattering
length $a^+$.  The main systematic error in  Eq. (\ref{eq:a+ determined}) comes
from the uncertainty in the dispersive correction term and, to a lesser degree,
from the form factor or non-locality in the deuteron double scattering term. 
The small corrections for isospin violation in the double scattering term and
for charge symmetry breaking in the single scattering on the deuteron are well
within the stated uncertainties and have no substantial influence on the
result. 

Combining the information from the experimental $\pi ^-$p and $\pi ^-$d 
scattering lengths with the constraints of the theoretical analysis  
(\ref{eq:a+ determined}), we obtain a substantially improved determination in
the difference $(a_{\pi ^-{\rm p}}-a_{\pi ^-{\rm n}})/2$ (this quantity is, in
the limit of isospin symmetry, identical to the isovector scattering length
$a^-$):

\begin {equation}
\frac {a_{\pi ^-{\rm p}}-a_{\pi ^-{\rm n}}}{2}=(895\pm 3{\rm (statistical)} \pm
13 {\rm (systematic)})\times
10^{-4}\ m_{\pi}^{-1}.
\label{eq:a- determined}
\end {equation}

\begin{figure}
\epsfig{file=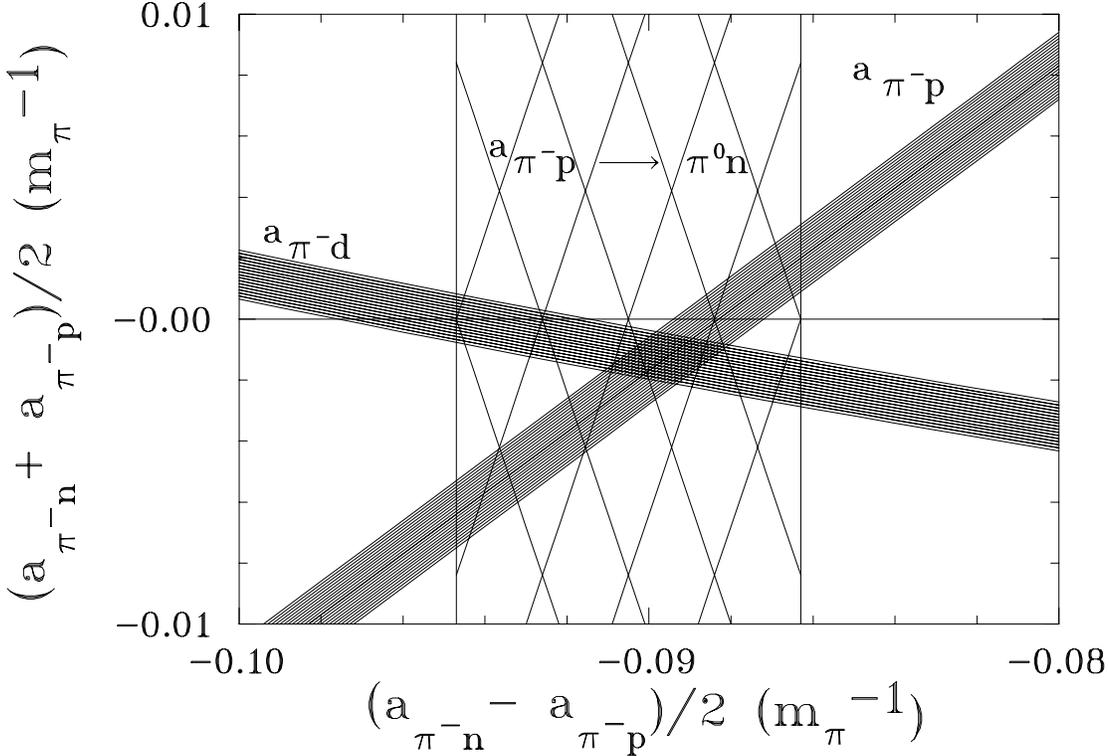,angle=90,height=10cm}
\caption{Graphical determination of the $\pi $N scattering lengths 
$(a_{\pi ^-p}+a_{\pi ^-n})/2 \simeq a^+$ and $(a_{\pi ^-p}-a_{\pi ^-n})/2 \simeq a^-$ from the constraints imposed by the pionic atom 
scattering lengths.}
\label{fig:aplus2609.ps}
\end{figure}

A graphical determination of these $\pi$N scattering lengths is shown in
Fig.~\ref{fig:aplus2609.ps}, which also emphasizes that this is a
substantial improvement on determinations using only data from pionic
hydrogen.  The results are in excellent agreement with the central values
deduced from the pionic hydrogen shift and width by the experimental PSI
group, since it follows from Eqs.  (7) and (8) of Ref.\cite {SCH99} that
$a^+=(-22\pm 43)\times 10^{-4}\ m_{\pi}^{-1};~~a^-=(905\pm 42)\times
10^{-4}\ m_{\pi}^{-1}$.  The PSI group
   \footnote
{After the submission of the present paper, the PSI group has published a
revised analysis\cite {SCH01} based on the B-K treatment\cite{BAR97a} and assuming strict isospin
symmetry. 
They quote
$$ a^+\equiv b_0=-0.0001^{+0.0009}_{-0.0021}m^{-1}_{\pi }
;\ a^-\equiv -b_1=0.0885^{+0.0010}_{-0.0021 }m^{-1}_{\pi }.$$ 
Their systematic errors are  not well controlled. First,  B-K explicitly omit the large dispersive correction, which contributes a term of the order of $ 0.0030 m^{-1}_{\pi
}$ to $ a^+$.  Second,
the sp interference contribution is negligible as we discuss in detail in
subsection IV.G, while it is derived in B-K from an erroneous assumption
with a value similar to that of the dispersive correction. The statement based on
their Ref.  [55] that the sp interference term partly could contain part of
the absorption term is incorrect.  In addition, the dominant contribution to their theoretical error appears to be based on a confusion about the form factor correction. They  introduce twice the B-K form factor effect,  counting it  as well as an (inexistent) off-energy shell correction of the deuteron wave function. This leads  to an overestimate of the lower systematic uncertainty from this source (double-counting).}
  also used the constraint from the
pionic deuterium shift, assuming the old calculation of Ref. \cite {THO80}
to be accurate enough and found $a^+=(+16\pm 13)\times 10^{-4}\
m_{\pi}^{-1};~~a^-=(868\pm 14)\times 10^{-4}\ m_{\pi}^{-1}$.

 From our evaluation here, we have achieved quantitative
control of the dominant contribution to the GMO relation from the scattering
lengths to about 1\% or better in $g_c^2/4\pi$.

It is interesting to compare our results with the extrapolations of scattering
amplitudes to threshold as given in Refs.  \cite{FET97,MAT97}.  They find the
value  $a_{\pi ^+{\rm p}\rightarrow \pi ^+{\rm p}} =(-770 \pm 30)\times 10^{-4}
\ m_{\pi}^{-1}$.  This corresponds to $a^+=(57\pm 15)\times 10^{-4}\
m_{\pi}^{-1}$ assuming isospin symmetry invariance and using the experimental
value for $a_{\pi ^-{\rm p}\rightarrow \pi ^-{\rm p}}$ from pionic hydrogen. 
On the other hand, the charge symmetric scattering length $a_{\pi
^-n\rightarrow \pi ^-{\rm n}}=(-917\pm 18)\times 10^{-4} \ m_{\pi}^{-1}$
follows from Eqs. (\ref{eq:a+ determined}) and  (\ref{eq:a- determined}) and
within charge symmetry the  two  values should be identical. According to Eq.
(\ref {eq:pi-p pi+p charge symmetry violation}) the estimated effect of charge
symmetry breaking in effective chiral theory is  $a_{\pi ^+{\rm p}}-a_{\pi
^-{\rm n}}=3\times 10^{-4}\ m_{\pi}^{-1}$. The above values give, instead,
$(147 \pm 35)\times 10^{-4}\ m_{\pi}^{-1}$, 50 times larger  than the expected
value.  Thus, unless charge symmetry is  unexpectedly badly broken, the
scattering length of  Refs.~\cite{FET97,MAT97} based on scattering experiments
is implausible and should be rejected.

While the extrapolation \cite{FET97,MAT97} leads to important differences,
it cannot, of course, be completely ruled out that other, more constrained,
extrapolations from $\pi $N scattering data could lead to scattering lengths
slightly different from the ones found here. The origin would then most
likely be due either to isospin violation in the scattering data or,
alternatively, to some unexpected modification of the least controlled part
of our deuteron terms, such as the absorption contribution. In the
dispersion-relation-constrained extrapolation advocated by Pavan {\it et
al.}~\cite {PAV99} they give $a^+=+20\times 10^{-4}\ m_{\pi}^{-1}$ to be
compared with 
$(-12\pm 8)\times 10^{-4}\ m_{\pi}^{-1}$
 above. Interpreted
as a modification of the dispersive term due to deuteron absorption, it
would require an increase by a factor of   2
in this term in order to make
the results compatible, which appears an implausibly large modification. We
believe our result to be the preferable one, since it is a more direct
determination and fully consistent.  The margin for modifications of our
theoretical analysis is small.

\section {Evaluation of the cross section integral $J^-$ from data }
\label {Evaluation of the cross section integral J^- from data.}

The  cross section  integral represents  only one  third of  the total
contribution to  the GMO relation.  This means that an  uncertainty of
(say)  3\% in  the integral  would give  only 1\%  uncertainty  in the
coupling constant.   At the  present precision, and  in spite  of this
insensitivity, this has  now become  one of
the  main sources   of  uncertainty in  the  determination of  the
coupling constant.  Since total cross  sections tend to  be inherently
accurate, the evaluation can be  performed with precision, but for the
high-energy region. There exists a vast amount of high quality data up
to very high  energies (beyond 240 GeV/c) and,  in the dominant region
below  1   GeV/c,  there  are  detailed  results   from  partial  wave
analyses.  The only previous evaluation with a
detailed discussion and  clearly stated sources of errors  known to us
is an unpublished study of  1985 by Koch, which gives $J^-=-1.077(47)$
mb~\cite {KOC85}.   Later evaluations find values within  this band of
errors, but the uncertainties are  not discussed. In 1992 Workman {\it
et  al.}~\cite {WOR92}  gave the  values $-1.056$  mb and  $-1.072$ mb
based on  the Karlsruhe--Helsinki  and VPI $\pi  $N amplitudes  of the
time,  respectively. In  1995 the  VPI  group gave  the value  $-1.05$
mb~\cite  {ARN95}.    Gibbs  {\it  et  al.}  give   a  similar  value,
$J^-=-1.051$~mb~\cite {GIB98}. In  this case the dominant contribution
below 2  GeV ($-1.308$  mb) was evaluated  using the  SM95 phase-shift
analysis~\cite{ARN95b} for the $\pi$N cross sections. These values are
summarized in Table V.

\begin{figure}
\epsfig{file=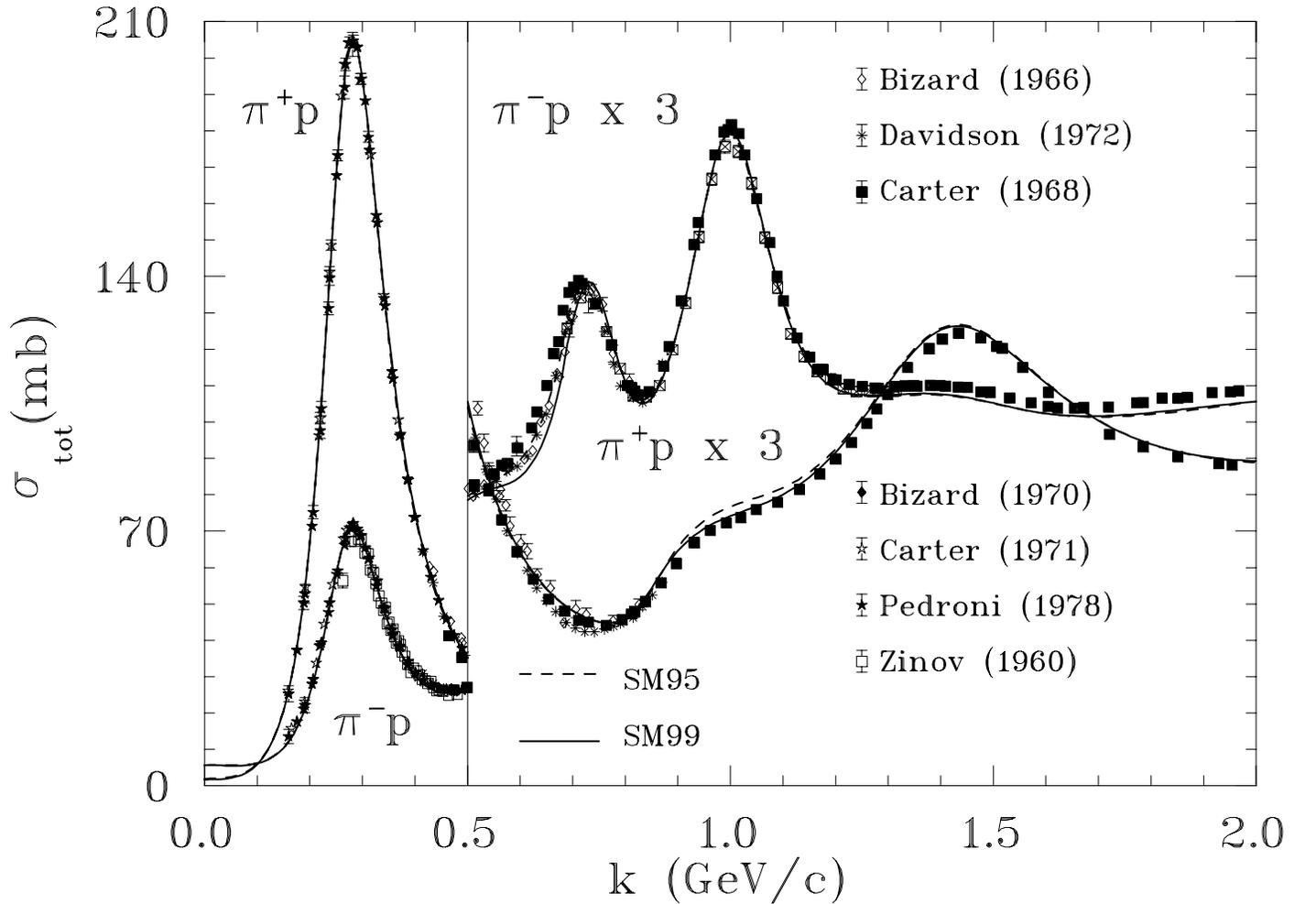,angle=90,height=13cm}
\caption{The experimental total $\pi^+$p and $\pi^-$p cross sections 
below 2 GeV/c~\protect\cite{BIZ66,DAV72,CAR68,BIZ70,CROSSKOCH,PED78,ZIN58}
 compared with the SM95~\protect\cite{ARN95b} and SM99~\protect\cite{ARN99}
 PWA hadronic solutions ,
 where 
Coulomb barrier effects have not  been taken into account.}
\label{fig:sigmatotal.ps}
\end{figure}

\begin{figure}
\epsfig{file=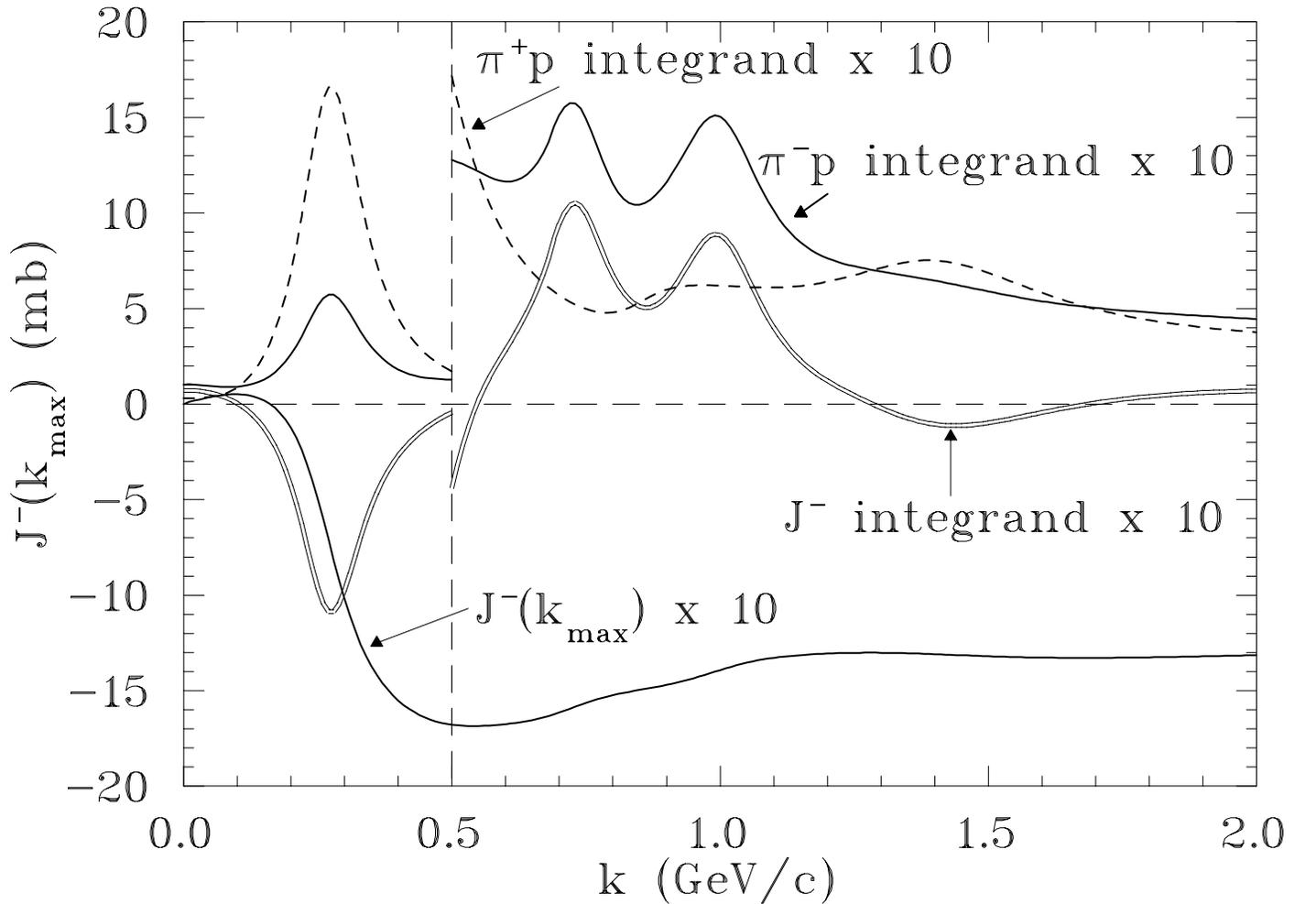,angle=90,height=13cm}
\caption{The separate integrands for $\pi^ {\pm}$p, as well as for
their difference as a function of laboratory momentum  $k$, 
together with the cumulative value of the integral $J^-(k_{{\rm max}})$ 
integrated from threshold to $k=k_{{\rm max}}$.  The curves are based 
on the SM99 solution~\protect\cite{ARN99}. The integrands are in units 
of mb GeV/c.}
\label{fig:J-integrand.ps}
\end{figure}

In view of  the importance of obtaining a clear  picture of the origin
of present uncertainties, we have re-examined this problem in spite of
the approximate consensus.   The $\pi $N total cross  sections below 2
GeV/c \cite{BIZ66,DAV72,CAR68,BIZ70,CROSSKOCH,PED78,ZIN58} are shown in Fig.  \ref{fig:sigmatotal.ps} and compared with the SM95\cite{ARN95b} and SM99\cite{ARN99} PWA hadronic solutions. The typical shape of
the  integrand $J^-$ is  seen in  Fig. \ref  {fig:J-integrand.ps}.  As
might be expected, the main  contributions come from the region of the
$\Delta $  resonance and just above.   It would be  false, however, to
believe that this is the  region that produces the main uncertainty of
the  integral. There  are  no strong  cancellations  in the  difference
between the total $\pi^{\pm }$p  cross sections in that region and the
cross  sections   have  been  very   carefully  analyzed.   Systematic
uncertainties contributing 2--3\%  or more to the total J$^-$
 are very  unlikely indeed;  if they
occur,  they  will certainly  have  an  important  influence on  other
determinations of the coupling constant as well.

In the following  we examine in detail the  uncertainties arising from
various   energy  regions   with  different   characteristics  (the
numerical conclusions  are summarized in  Table VI and VII): 
\newline 
 - in
subsection A, the threshold region below 160 MeV/c is dominated by the
s- and p-wave threshold parameters (s-wave contribution of about +6\%,
p-wave one of about $-6$\%);  \newline - in subsection B, the $\Delta$
resonance region from 160 MeV/c to 550 MeV/c, in which the major phase
shifts are  very accurately known (main contribution  of about 155\%);
\newline - in  subsection C, the resonance region from  550 MeV/c to 2
GeV/c, which is partly dominated by higher resonances with mostly high
quality data (about $-33$\% contribution); 
\newline - in subsection D,
the  high-energy region  and the  asymptotic  region from  2 GeV/c  to
$\infty$ (totally  about $-22$\% contribution);  about half originates
from  the  asymptotic region  beyond  10  GeV/c,  for which  data  are
accurately described by asymptotic expressions.

The total cross sections in the integral $J^ -$ are the hadronic ones. The
experimentally defined total cross sections differ from these due to the
electromagnetic corrections. These are nearly model independent in the
present context. They are proportionally more important in the difference
between the cross sections, since the $\pi ^+$p total cross sections are
systematically reduced at all energies by the Coulomb repulsion between the
particles and, conversely, the $\pi^-$p ones are systematically increased by
the attraction\cite{BUG74,BUG99}. This effect gives a positive contribution
to $J^-$; the coupling constant would be underestimated by about 3\%
neglecting such corrections. The dominant correction comes from the $\Delta$
resonance region (see Table VI). For total cross sections there is little
sensitivity to the detailed procedure: the Nordita approach is frequently
used\cite{TRO78} below 500 MeV/c. The Coulomb correction to the integrand at
high energy, where the $\pi ^{\pm }$ total cross sections are nearly equal,
is approximately $(4\pi ^2)^{-1}2A_c\sigma ^T(k)/k^2$ with $A_c \simeq 3.7$
MeV/c.  For constant cross sections, the total correction above a momentum
k$_1$ is then typically 0.007$k_1 ^{-1} $ mb, where k$_1$ is in units of
GeV/c\cite{BUG74}. It therefore rapidly becomes negligible above a few
GeV/c.

\begin{figure}
\epsfig{file=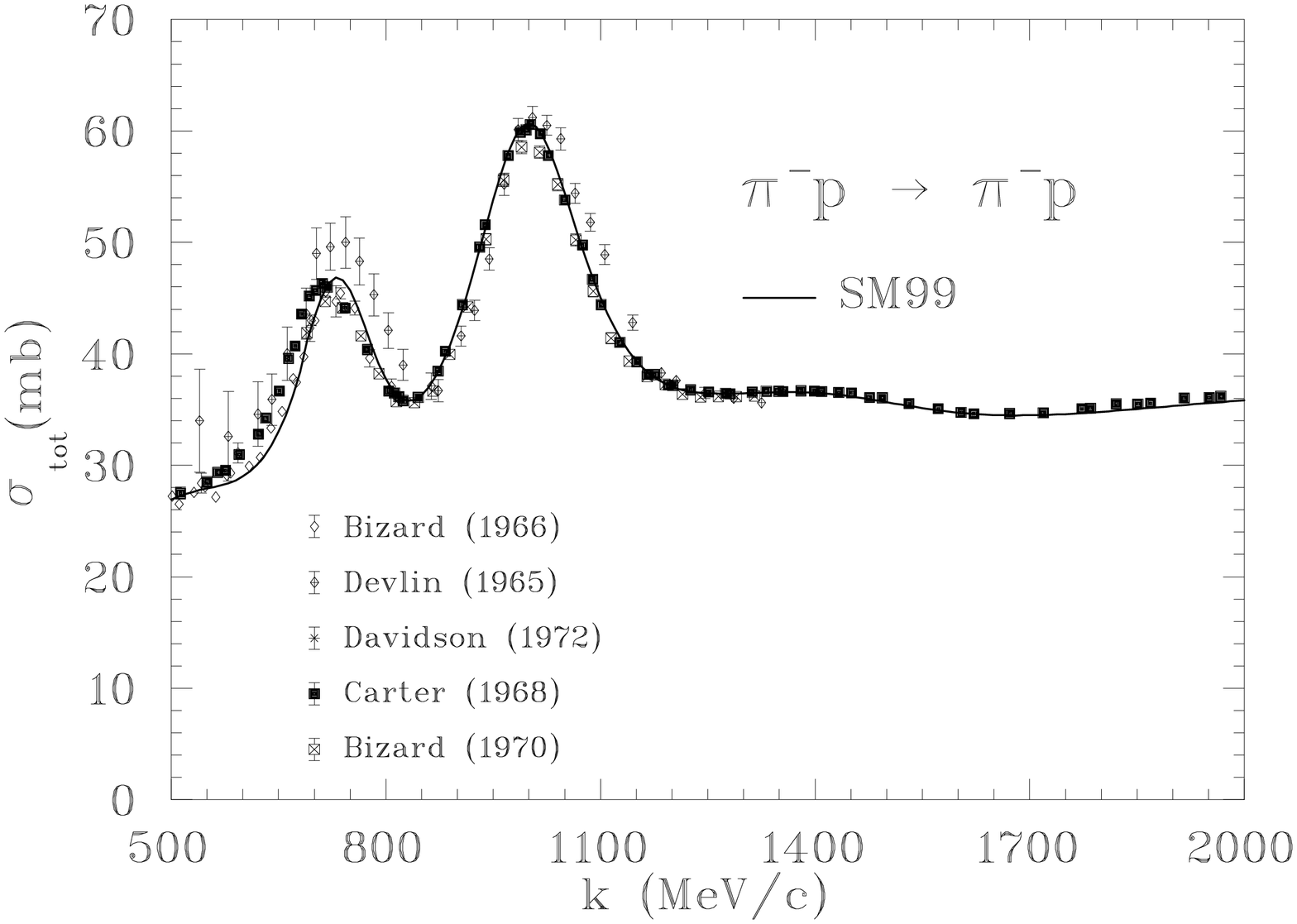,angle=90,height=12cm}
\caption{The $\pi ^-$p experimental total cross sections in the region
$0.5\le k\le 2$~GeV/c~\protect\cite{BIZ66,DAV72,CAR68,BIZ70,DEV65}
  compared to SM99~\protect\cite{ARN99} with Coulomb 
barrier effects accounted for.}
\label{fig:pimptot04.ps}
\end{figure}

\begin{figure}
\epsfig{file=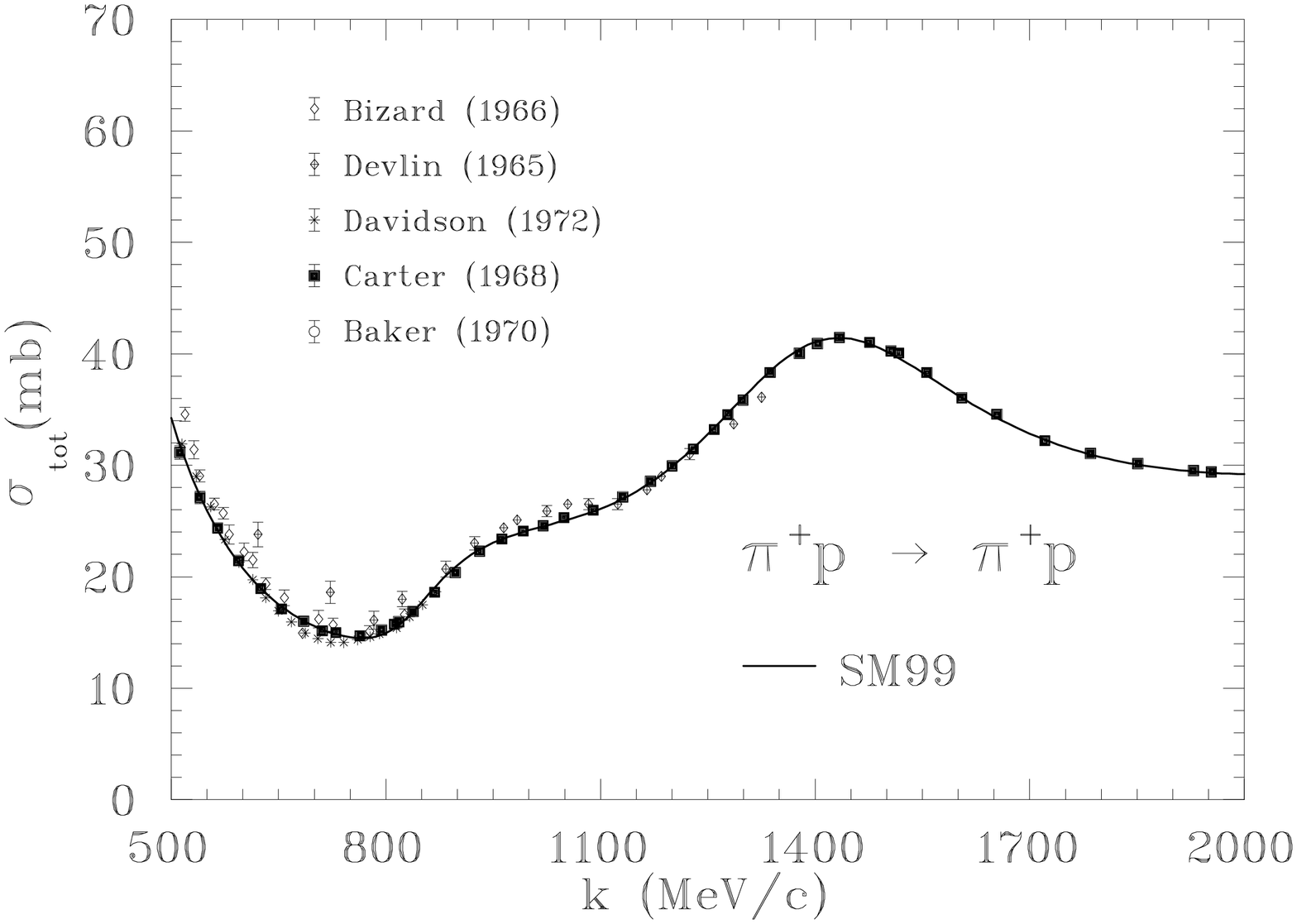,angle=90,height=12cm}
\caption{The $\pi ^+$p experimental total cross sections in the region
$0.5\le k\le 2$~GeV/c~\protect\cite{BIZ66,DAV72,CAR68,DEV65,BAK70}
  compared to SM99~\protect\cite{ARN99} with Coulomb barrier effects
accounted for.}
\label{fig:pipptot04.ps}
\end{figure}

 As  an illustration of contributions, the resulting fits to data\cite{BIZ66,DAV72,CAR68,BIZ70,DEV65,BAK70} for
the solution  SM99 of
 Arndt {\it et  al.} are shown in  the range $0.5
\le  k_{lab}\le  2$~GeV/c  in  Figs. \ref{fig:pimptot04.ps}  and  \ref
{fig:pipptot04.ps}.

 The recent VPI/GWU partial wave amplitude (PWA) solution up to 2
 GeV/c\cite{ARN99} is in good agreement with observations with a few
 exceptions.  We will therefore use the hadronic cross sections deduced from
 this solution as a guide for the numerical contribution.  We estimate its
 uncertainties below.  We also give numbers from the earlier PWA solution
 SM95~\cite{ARN95b} for comparison.

\subsection { The threshold region} 

There are no direct measurements of total cross sections below 160 MeV/c, but
the  hadronic  cross-section difference  can be well reconstructed from other
considerations. In this range the  low-energy  s- and
p-wave parameters determine the result.  The cross-section difference at
threshold is 

\begin {equation}
\sigma^T_{\pi ^-{\rm p}}(0)-\sigma^T_{\pi ^+{\rm p}}(0)=
8\pi \ \left((a_{\pi ^-{\rm p}})^2-(a^{+})^2\right),
\label{eq:threshold cross section}
\end{equation}
assuming isospin invariance and neglecting the mass differences. Here the first
term is accurately known from the $\pi ^-$p atom, as previously discussed, and
the second term is extremely small. With increasing energy the p-wave
contributions of opposite sign, governed by the tail of  the $\Delta $
resonance, take over and compensate the s-wave term beyond 100~MeV/c. These two
terms contribute together +0.011 mb~\cite {ARN99}, but taken individually the
s- and p-wave terms represent about 6\% each of the total $J^-$. The
uncertainty is dominated by the error in the rather small contribution from the
s-wave range terms, while the accurate $\pi ^-$p scattering  length is imposed
in the SM99 analysis. The corresponding error in $J^-$, of about  0.5\%, is not
a major source of overall uncertainty and even if this uncertainty is
underestimated this has little importance.

The 3.3 MeV mass difference in the $\pi ^-$p and $\pi ^0$n thresholds breaks
the isospin invariance leading to a potentially significant correction, in
particular, since the $\pi ^-$p total cross section diverges at threshold due
to the open $\pi ^0$n channel.  The smallness of the contributions from the
threshold region hints at a small correction. We have investigated this effect
using a simplified model based on the s- and p-wave  low-energy  parameters
only. The dispersion relation must now be evaluated using the imaginary part of
the scattering amplitude Im$F = 4\pi k\sigma ^T $,  which is well behaved at
threshold, but which differs from zero below the physical $\pi ^-$p threshold.
The correction occurs predominantly in the 6\% s-wave term. The approximate
modification up to the momentum $k_1$=160 MeV/c is of O(-$\kappa
^2/2k_1^2)\simeq -0.02$, where $\kappa ^2 
\simeq 0.045 \ m_{\pi }^2 $ is the  $\pi ^0 $ squared momentum at threshold.
This represents a  $-0.1 \%$ contribution to the integral $J^-$, which  is
negligible compared with other uncertainties.

\subsection { The $\Delta $ resonance region} 

This  is  the  main  contribution  to  the integral  and  it  must  be
accurately evaluated.  The resonant  33 wave dominates heavily and its
behavior is strongly constrained  by other experiments and theory. The
main contribution  comes from  the $\pi ^+$p  cross section,  which is
approximately  three  times  larger   than  the  $\pi  ^-$p  one. 

The  systematic uncertainties  are  more  important than  the
statistical ones.  In order to judge their importance,
we first  evaluated this  contribution directly from  the experimental
$\pi ^+$  and $\pi  ^-$ data sets  taken separately,  with statistical
errors added  in quadrature  \cite{PDG98}.  The result  is $-1.652(6)$
mb.   It differs  by only  0.020  mb from  the corresponding  quantity
evaluated from the phase  shift solution SM99.  Since the total
cross  sections  have incoherent  sums  of  the  squared partial  wave
amplitudes  the large,  accurately known,  phase shifts  dominate.  The
phase  shift solution incorporates  strong additional  constraints and
eliminates minor inconsistencies in the  data and is preferable to the
raw data.  The e.m. corrections to J$^-$ come mainly from this region.
They are  only weakly model dependent  and are included  in SM99 using
the Nordita  procedure \cite{TRO78}. This correction  amounts to 0.060
mb as seen in Table VI  for the difference in J$^-$ evaluated hadronic
vs.  nuclear cross sections from  the same phase shift solution.  This
well controlled correction increases the coupling constant
by only 1.8 \% and it represents the main e.m. correction.

 A modern analysis such as SM99 favors the use of the experimental cross
sections dominated by the data of Pedroni {\it et al.}~\cite {PED78}; the
cross sections from Carter {\it et al.}~\cite {CROSSKOCH}, which dominated
the analysis in the 1980's, would lead to a more negative value for $J^- $
and, correspondingly, to a $\pi $NN coupling constant larger by about 1\%.
The hadronic SM99 total cross sections do not contain the inverse
photo-production cross section, which contributes 1 mb (1.5 \%) of the total
nuclear $\pi ^-$p cross section at the resonance peak. This is a negligible
source of uncertainty.

 The $\Delta $ mass splitting may also affect the coupling constant deduced
from determinations based on $\pi$N data.  The empirical isovector mass
splitting directly observed in the null experiment by Pedroni {\it et
al.}~\cite {PED78} corresponds to $M_{\Delta ^0}- M_{\Delta ^+} =1.38(6) $
MeV. To our knowledge, there exists no information on the isotensor
splitting, and we neglect it. With respect to strict isospin symmetry and
with the effective position of the $\Delta $ resonance that of the $\Delta
^{+}$, the correction to J$^-$ is approximately $4(M_{\Delta ^+}- M_{\Delta
^0})/3(M_{\Delta }-M)\simeq -0.6\% $, which would {\it increase} the
coupling constant $g_c^2/4\pi $ by 0.2\%. The same conclusion follows from a
study by Arndt\cite{ARN99a} who used a mass splitting of 0.2 MeV and the same effective
$\Delta $ position. The corresponding change in J$^-$ is only 0.05\%, which
scaled to the observed mass splitting contributes 0.17 \% to g$^2/4\pi $.
Consequently, the mass splitting does not substantially influence the value
of g$^2/4\pi $.

\subsection { The resonance region} 

This region from 550 MeV/c to 2  GeV/c  is, as a whole, well measured and is
analyzed in SM99. The contributions to the integral are positive and rather
important up to 1.2  GeV/c, partly compensating the  contribution from the 
$\Delta $ resonance region. The remaining region contributes little, but is a
minor source of uncertainties.

It is interesting to quantify the difference between a state of the art
phase shift solution and data in more detail (see Table VI). In the region
of 550~MeV/c to 1.2~GeV/c the nuclear SM99 solution gives $0.012$~mb less
contribution to $J^-$ ($-1.1$\%) than the direct experimental cross
sections, while from 1.2 to 2~GeV/c the corresponding contribution is
$0.008$~mb less ($-0.7$\%).  The overall e.m.  corrections to the integral
in this region 0.041~mb or 1.3 \% in g$^2/4\pi $ (see Table VI). The SM99
solution assumes a point charge distribution. Improved Coulomb corrections
using an extended charge distribution\cite {GIB01}
are being implemented in the phase shift solution SP02, but the results are
little changed \cite {PAV01}. Assuming pessimistically that the correction
is accurate only to 33\%, the overall uncertainty from this source would
still be only $\pm $0.014 mb in $J^-$ or $\pm 0.4\%$ in the coupling
constant. The main uncertainty is therefore not due to the e.m. correction,
but to systematic differences between SM99 and data. It comes mainly from
the region just above 550 MeV/c, as will now be discussed.

At the low energy end of the region between 550 and 1200 MeV/c, there are
long-standing experimental problems of systematic nature with the total cross
section data. Those of Davidson {\it et al.}~\cite {DAV72} have an incorrect
energy calibration, too low by about 10 MeV/c, and its 72 data points must
either be re-calibrated or eliminated from the analysis
\cite{PAV99,HOE83,PAV99a}.  Similarly, the SM99 solution, driven by modern
angular distributions, is systematically lower than the $\pi^-$p data of
Carter {\it et al.}~\cite{CAR68} below 700 MeV/c, a region where data for
experimental reasons, are less reliable than at higher energies. These
points have been omitted from the PWA analysis\cite{PAV99,PAV99a} (see also
Fig. \ref{fig:pimptot04.ps}).  This discrepancy is larger than the e.m.
corrections in the same energy region.  Under the circumstances we have
preferred to use the SM99 PWA solution as the best guide, but we use the
difference with data as a liberal measure of the uncertainty.  We therefore
use the overall SM99 contribution from this region of $0.378~ \pm $
0.020~mb.

\subsection { The high-energy and asymptotic regions}

There exists abundant experimental information on $\pi ^{\pm}$p cross
sections to high precision from 2~GeV/c up to 350~GeV/c. The main
uncertainty in $J^-$ in this region is associated with the relatively slow
convergence of the integral. At energies beyond 4~GeV/c there has been an
important effort to measure and analyze cross sections, since the issue of
the rate at which the $\pi ^{\pm}$p cross sections become asymptotically
equal, is important theoretically for asymptotic theorems. The discussion below is
summarized in Table~VII.

The region $2\le k\le 4.03$ GeV/c has been  calculated
using the Particle Data Group (PDG) 1998 tables~\cite{PDG98} (see also \cite{CAR68,CIT66}) and gives a
moderate contribution of 0.064~mb, with a modest $\pm 0.007$~mb systematic
error. The 
statistical uncertainty is small. Beyond this region, cross section data with
considerable systematic and statistical accuracy exist from $4.03~ \le k\le
370$~GeV/c and are listed in the PDG tables \cite{PDG98,PDG94}.  We first
evaluated the contribution directly from the precision data. This gives
0.133~mb in the range $4.03 \le k\le 240$~GeV/c, with a small statistical
error and a systematic error of about $\pm 0.022$~mb or $\pm 1.8$\% in $J
^-$.
In addition, the 1994 version of the PDG tables\cite {PDG94} also lists a fit
to these data from 4.03 GeV/c to 240 GeV/c (Table 33.3).  Using the fitted expression, we
have evaluated the contribution in the same interval as above using this
expression and find 0.155~mb. This is 0.022~mb  higher than the value of
0.133~mb  by direct evaluation, but in good general agreement. This larger
value has been used in several previous GMO evaluations\cite {ARN94a,GIB98}.
 We prefer   the lower value as
 more transparently linked to the actual data. 

Finally, there is a small, but not negligible, contribution from the very
high-energy region from 240 GeV/c to $ \infty $.  We determine this from the
Donnachie--Landshoff Regge fit to the data~\cite {DON92}, which describes the
observed cross section difference well at the highest energies. This fit is a
sum of two Regge terms, one arising from Pomeron exchange and the second from
lower-lying resonance exchange.  It gives a contribution of 0.030~mb.
Alternatively, one might consider using the three-term fit (one for the Pomeron
and two for the Reggeons) in the 1998 PDG tables (Table~38.2), which gives
0.018~mb. This low value is not surprising, since the 1998 PDG parameterization
gives a difference 27\% lower than  the PDG 1994 one in the region above
200~GeV/c, at variance with the data~\cite {CARR80}. At lower energies, this
parameterization agrees better with the  data in the region of 100~GeV/c~\cite
{CARR80,CARR74}. We have also used a recent high-energy fit based on a
two-Pomeron pole expression fully  compatible with universality, Regge
factorization, weak Regge exchange degeneracy,  and  generalized Vector
Dominance Model~\cite{GAU00}.  This parameterization (see Eq.~(13) and Table~1
of Ref.~\cite{GAU00}) gives a contribution of  0.025~mb. The corresponding
uncertainties, given in  Table~VII, come mainly from the 4\% uncertainty in the
Regge intercept. This spread of values according to the model considered for
the fit introduces  an additional systematic uncertainty of 0.006~mb from this
high-energy  region.
 The integrated  Coulomb correction above 2 GeV/c is negligible, since it
 is only 0.003 mb
using  the   estimate  of  Ref.  \cite{BUG74} as given
 in the beginning of this section. 

\subsection
{Summary of the results for $J^-$}

  The purpose of this section has been to establish the importance
 of different energy regions for the integral J$^-$ and their 
contribution to the uncertainty. Since there exists total cross sections
 from state of the art partial wave analysis up to 2 GeV/c we first
 studied the accuracy to which such an approach describes contributions
 to  J$^-$ based on actual data in the region 0.16 to 2 GeV/c. 
To this end, we  evaluated the contributions to  $J^-$ from data in
 different energy regions with  no
Coulomb corrections  other than  those introduced by  the experimental
authors (nuclear cross sections). The
results are  summarized in Tables VI and VII.  The  statistical uncertainty in
the  evaluation is  small. The  trapezoidal formula  was used to
integrate the data and  the corresponding statistical errors were
added quadratically. Within the different integration ranges, as given
in the Tables, the systematic  error was calculated by varying the
experimental  results  inside  the  interval  defined  by  the  quoted
systematic  error. The full systematic  uncertainty was  obtained by
the  quadratic sum  of the  error in  each interval, since their origin
 is different. We then confronted these results with the analogous quantities
obtained from the partial wave analysis SM99. The main deviations occurs in 
the experimentally difficult region 0.55-0.70 GeV/c. Since the partial 
wave solution incorporates additional experimental constraints, we consider 
it superior to the direct data in the crucial region and we base the further 
analysis on the PWA solution SM99.  

We then examine the e.m. corrections.  These are under theoretical control
inside the PWA analysis. The uncertainties in these corrections are less
 important than the systematic difference between data and the PWA solution.
 The  consequence  of   the  $\Delta  $  mass  splitting  is
negligible. 

The low energy region below 0.16 GeV/c contributes little to J$^-$ and there are
no experimental total cross sections in this region. It is 
strongly constrained by the $\pi ^-$p scattering length and
 the tail of the $ \Delta $ resonance such that it can be well 
controlled without the necessity of e.m. corrections. 

 The systematic uncertainty has
its origin principally in the region above 4 GeV/c.  There is also a sizeable
systematic uncertainty that is due to the inconsistencies of the $\pi ^-$p data
in the region 550-700 MeV/c, although we have probably overestimated this
uncertainty. We find from Table VII, rows 15--17, that three
different descriptions, based on the SM99 PWA below 2~GeV/c,  give values in a
rather narrow range; $-1.087\pm 0.009\pm 0.031$ ~mb, $-1.099\pm 0.008\pm
0.031$~mb and $-1.092\pm 0.009\pm 0.031$~mb. The difference between these
values is smaller than the estimated systematic uncertainty.  We also give in
row 14 the less negative result obtained with the older SM95 PWA below 2~GeV/c
and the fit PDG94 in the momentum range from 4.03 to 240 GeV/c: $J^-= -1.053\pm
0.010\pm 0.031$~mb. We have chosen the average of these four values as
characteristic of the integral. The systematic uncertainty provides an
adequate  band of possible values, so that

\begin {equation}
 J^-=  -1.083\pm 0.009\pm 0.031 ~{\rm mb}.
\label{eq:Jminusvalue}
\end{equation}

Our result for  $J^-$ is close to the unpublished  value of Koch \cite
{KOC85},  J $^-=  -1.077\pm 0.047  $ mb,  which is  the  only previous
explicitly documented  and detailed evaluation  known to us.   The  main
difference in the input data with Koch is an updated evaluation of the
contributions from the high-energy region and better control of e.m.
corrections.  We show also that the $\Delta $ mass splitting is unimportant
and include an improved  discussion of the threshold region using
modern data. 
 It is important to
realize that  the main uncertainty to  J$^-$ comes from  the very high
energy contribution. It is difficult to ascribe a major uncertainty to
  the Coulomb corrections.   We
note that the previous evaluations  of J$^-$ quoted in Table V without
uncertainties stay within our range of errors.

\section {Results}
\label {Results}

In conclusion, we summarize our work as follows. We first derived new values
for the $\pi $N scattering lengths, using the $\pi ^-$d atomic data analyzed
in an improved theoretical approach. The statistical and systematic
uncertainties contributions were thoroughly examined. The corresponding $\pi
^-$d scattering length gives a nearly direct determination of the small
'isoscalar' $\pi ^-$N scattering length to good precision. From this
constraint together with the $\pi ^- $p scattering length from pionic
hydrogen we obtain a high precision value also for the isovector length. In
fact when we examine the basic experimental input of the highly accurately
quoted scattering length $a_{\pi ^- {\rm p}}$, deduced from the $\pi ^-$p
atomic energy shift and quoted to high accuracy~\cite {SIG96}, we found that
there are small inconsistencies in their current procedure at the level of
$\pm 1\%$.  This should be improved, since the precision is otherwise
unsatisfactory for the determination of the $\pi$NN coupling constant.  In
addition, the experimental accuracy is now so high that systematics in the
theoretical analysis of the $\pi ^-$d scattering length is the main source
of uncertainty in the disentangling of the isospin components of the $\pi $N
scattering length. The dominant limitation to higher accuracy is the
dispersive contribution from the physical absorption process $\pi ^-$d
$\rightarrow $ nn.  A thorough modern re-examination of this contribution is
highly desirable. Our analysis does not assume strict isospin symmetry,
although we do not see any signs of violation at the present level of
precision. We present the results, however, so that they can be directly
used in discussions of the validity of this symmetry. The values we find
using the empirical $\pi ^-$p and $\pi^-$d scattering lengths from
subsection 4.1, Eqs. (20) and (21), are

\begin {equation}
a^+\simeq \frac {a_{\pi ^-{\rm p}}+a_{\pi ^-{\rm n}}}{2}=
(-12\pm 2 \pm 8)\times 10^{-4}\ m_{\pi}^{-1}.
\label{a+final}
\end {equation}  
\begin {equation}
a^-\simeq  \frac {a_{\pi ^-{\rm p}}-a_{\pi ^-{\rm n}}}{2}= 
(895 \pm 3 \pm 13)\times 10^{-4}\ m_{\pi}^{-1}.
\label{a-final}
\end {equation}

These values are based on  theoretical improvements on previous
work.  (See also comments and footnote after Eq. (21).)

Our second conclusion concerns the charged $\pi $NN coupling constant, which
can be derived from the GMO forward dispersion relation, using our new,
accurate value for the symmetric $\pi $N
scattering length (\ref{a+final}). 
Use of Eq. (\ref
{eq:GMOrobust}) assuming charge symmetry and with input values from Eqs.
 (\ref {eq:exp pi-p scattering length}, \ref {a+final}) as
well as with $J^-=(-1.083\pm 0.009 \pm 0.031)$ from Table V  gives :

\begin{equation} g_c^{2}/4\pi = (4.87 \pm 0.04 \pm 0.14) +
(9.12 \pm 0.02 \pm 0.10) + (0.12 \pm 0.02 \pm 0.08)  
=(14.11\pm 0.05 \pm 0.19).
\label{g^2final}
\end {equation}
 The uncorrelated  statistical and systematic  uncertainties have been
added  separately in quadrature.   The main  uncertainty is  no longer
dominated  by the  scattering  lengths,  but comes  as  much from  the
weighted  integral $J^-$  of the  difference between  the charged-pion
total cross  sections. Its dominant systematic  uncertainty comes from
the  region above  4  GeV/c.  Previous  determinations  using the  GMO
relation \cite{WOR92,SCH99,GIB98,ARN95} will all give similar results,
provided one uses  the empirical scattering lengths, which  are by now
well established.  
 \footnote {After the  submission of our  paper, the
PSI group evaluated the  GMO relation from scattering lengths obtained
using the B-K corrections  to the pion-deuteron scattering length\cite
{SCH01}.  We have  discussed  the problems  of  this determination  in
Section IV.M, footnote 1. Their  quoted value $ g^2/4\pi = 13.89^{+0.23}_
{-0.11}  $  is  consistent  with  our  result,  but the systematic uncertainties are not well controlled. In particular, there  is  a
substantial additional  systematic error of  about 0.25 or more from the isoscalar scattering length.   In addition,
they use a value for $J^- $  derived from an average of those given
in Refs.  \cite  {WOR92,GIB98,ARN95} with the spread of  values as the
only uncertainty.  Of these values, two  (Refs.\cite {WOR92,ARN95}) do
not state any uncertainty at  all, while Ref. \cite{GIB98}  states
the (small)  statistical uncertainty  only. In particular,  the rather
large  uncertainty  from  the   high  energy  region  4-240  GeV/c  is
neglected.  Note that  the Particle Data Group 1994  fit to the region
4-240 GeV/c gives a +0.02 mb higher contribution to $J^- $ (Table VII,
line 10)  than the direct  data of the  1998 version (Table  VII, line
9).  The latter corresponds  to a  0.10 higher  value of  the coupling
constant.}

The value,  $ g_c^{2}/4\pi =14.11$,  which we obtain for  the coupling
constant is intermediate  between the low value of  about 13.6 deduced
from the large data banks of  NN and $\pi $N scattering data using the
PWA  approach  \cite  {STO97,TIM97,ARN94a}   and  the  high  value  of
14.52(26)  from np  charge exchange  cross  sections\cite{RAH98}.  The
uncertainties in the determination  of the coupling constant using any
method are dominated by  systematics.  Consequently, we have refrained
from combining our result  with those from other approaches.  However,
if  the systematic  error  were to  have  Gaussian distributions,  our
result  differs  from that  of  Uppsala  \cite  {RAH98} by  only  1.25
standard  deviations  (21\% probability)  and  from  that  of Pavan  et
al.\cite {PAV99} by 1.7 standard deviations (8\% probability). The PWA
results  have probably  systematic errors  far larger  than  the small
statistical errors to judge from the corresponding situation using the
data banks  with dispersive constraints\cite {PAV99}, but  this is not
quantified yet.  The modification  of the value  of $J^-$  required to
accommodate a  value of 13.6 is about  10\%. The major part  of such a
modification would  most likely  come from the  region above  2 GeV/c,
which implies  changes in  the contributions from  that region  of the
order of 50\%. Such large changes appear unlikely to us.

We therefore conclude that the present evaluation of the GMO sum rule, with
quantitatively controlled uncertainties in the input values for the $\pi $N
isoscalar scattering length, as well as for the cross section integral $J^-$,
does not readily support the conclusion of the indirect PWA determinations that
the $\pi $NN coupling is close to 13.6.  It should be noted that our
value has consistently been evaluated in a conservative way, such that the
parameters used in  the evaluation systematically lead to a value for the
coupling constant, which is  somewhat on the low side.

The strongest support for a relatively low value of the coupling constant comes
from the careful dispersive  analysis by  Pavan {\it et al.}~\cite {PAV99},
based on the VPI/GWU PWA description of $\pi $N scattering. It selectively
concentrates on pion-dominated amplitudes.  They find a value of  $13.73~\pm 
~0.01~\pm ~0.08 $,  where the first uncertainty is statistical and the second
systematic.  The authors use a variety of dispersive methods and find
$a^+=+0.0020\times 10^{-4} \ m_{\pi }^{-1}$. This value is small, but it has
the opposite sign from ours. They evaluate the GMO relation as a consistency
check and find a value of $g_c^2/4\pi=13.75$,  in agreement with their
dispersion result.  Since their evaluation is constrained by the experimental
$\pi ^-$p scattering length and their value for the dispersive  integral $J^-$
is nearly the same as ours, which is based to a large extent on their PWA
analysis, the difference with our result must be almost entirely ascribed to
the difference  in the value of $a^+$, a small  quantity,  which is difficult
to calculate from  scattering data. The origin of this difference is  not known
yet, but it might originate in the treatment of small electromagnetic
corrections to the scattering data. The minor  inconsistency in their analysis
is of little importance for most of their discussions, but it  becomes highly
relevant in the present context.

It is interesting to examine the consequences of our analysis for  the
Goldberger--Treiman (GT) discrepancy \cite {Gol58}. Following the discussion in
Ref.\cite {RAH98} the value for the coupling constant  found here  corresponds
to a discrepancy of 
$\Delta _ {{\rm GT}}=(3.6~\pm 1.0)\%  $,
 with $ \Delta_{{\rm GT}}$ 
defined as
\begin{equation}
 g_c\ (1-\Delta_{{\rm GT}}) ={\rm M}\ g_{{\rm A}}/f_{\pi }.
 \end{equation}

This corresponds  to a  $\pi $NN monopole  form factor with  a cut-off
$\Lambda ~=~ 800 ~\pm $ 80 MeV/c.  There exists no direct experimental
information on this form factor, which is inherently an off-mass-shell
quantity.   On the other  hand, within  the framework  of PCAC,  it is
naturally  expected to  be similar  to the  axial form  factor  of the
nucleon, a  dipole with a 1  GeV/c cut-off. This  expectation has been
confirmed  in  many  models,  using  a  variety  of  approaches  \cite
{GUI83,BOC99,LIU95,MEI86,KAI87},  beginning with  Ref.  \cite {GUI83}.
Such values  are fully consistent  with our findings for  the coupling
constant.  In contrast to these rather soft form factors, the deuteron
properties,  and  in  particular  its quadrupole  moment,  require  an
effective  cut-off  of 1.3  GeV/c  or  more,  since the  tensor  force
otherwise  becomes  too  weak~\cite  {MAC87,ERI83}.  It  is,  however,
believed at present that this  hard effective form factor is generated
by the correlated  exchange of an interacting $\pi  \rho $-pair, which
generates additional  tensor strength, when  explicitly accounted for:
the   true    one-pion-exchange   form   factor    is   softer   \cite
{HOL90,HAI92,JAN93,JAN94}.  A  low  value  for the  coupling  constant
should  therefore not  be  considered an  advantage  in resolving  the
Goldberger--Treiman discrepancy.

 Additional  support  for a  coupling  constant  $g^2/4\pi $  somewhat
 larger than 14 comes from  the recent measurements by Raichle {\it et
 al.} of polarized np  total cross sections~\cite {RAI99}. From these,
 the pion-dominated  $\epsilon _1$ parameter can  be determined.  They
 find  that  it  is  systematically  larger than  the  values  in  the
 phase-shift analysis  PWA93 of the  Nijmegen group~\cite{STO93a}.  If
 the discrepancy persists in other PWAs, this observation suggests, as
 a possible partial explanation, that the PWA coupling constant is too
 small.  In  any case,  it points to  an unexplained  discrepancy with
 those PWA analyses on which  the argument for a low coupling constant
 is based.

In order  to facilitate  future improvements on  the present  work, we
have   presented  the  various   corrections  in   such  a   way  that
modifications   of  any  individual   contributions  can   be  readily
incorporated without  the necessity of a complete  re-analysis. We see
three main  areas in  which the present  work can be  improved. First,
theoretical investigations of the relation between the hadronic energy
shift of the pionic atom and the scattering length should diminish the
present uncertainty in the deduced  $\pi ^-$p and $\pi ^-$d scattering
lengths by  a factor of  at least 2.  Second, the measurement  to high
precision of the width in  pionic hydrogen should give a separation of
the isospin components in the  $\pi ^-$p scattering lengths to similar
precision  as  that  obtained  from  the deuteron  data,  but  without
invoking deuteron  structure. Third,  studies of the  dispersion shift
for threshold pion absorption on the deuteron should eliminate a major
uncertainty in  the theoretical treatment of the  $\pi ^-$d scattering
length.  This  would  allow  the  $\pi $NN  coupling  constant  to  be
determined to 1\% precision.

\section {Acknowledgments}
\label {Acknowledgements}

We are particularly indebted to  R. L. Arndt, V. V. Baru, A. E. Kudryatsev, M. 
Pavan and M. Sainio for providing us with unpublished material and to D. Bugg,
M. Ericson, G.  F\"aldt, J. Gasser, W.~R.~Gibbs, G.~H\"ohler, O.~Krehl,
S.~Krewald, M. ~Lacombe, U.-G. Meissner and S. Wycech for valuable information.
This work has been supported by grants from the Swedish National Research 
Council and from the Swedish Royal Academy of Sciences.

\appendix
\section{Expressions used for the electromagnetic corrections to the 
experimental $\pi^-$d scattering length}

We here give details on expressions we used in the evaluation  of the 
electromagnetic corrections to the experimental $\pi^-$d scattering length.
One can write, with $q^2=\omega^2-m_\pi^2$, the low-energy $\pi$N 
amplitudes as,
\begin{eqnarray}
a_{\pi^-p}(\omega)&=&a^+ + a^- + (b^+ + b^-)\ q^2 \nonumber \\
a_{\pi^-n}(\omega)&=&a^+ - a^-  + (b^+ - b^-)\ q^2.
\end{eqnarray}
H\"ohler~\cite{HOE83} gives $b^-=(133 \pm 60) \times 10^{-4}\ m_{\pi}^{-3}$
and
$b^+=(-443 \pm 67) \times 10^{-4}\ m_{\pi}^{-3}$. 
The $\pi^-$d single scattering term is
\begin {equation}
S~=~\lambda _1 \ [a_{\pi ^-{\rm p}}(\omega )+ a_{\pi^-{\rm n}}(\omega )]=2
\lambda _1\ ( a^+ + b^+ q^2)
\end {equation}
where,
\begin {equation}
\lambda_{1}~= ~\frac {(1+m_{\pi}/M)}{(1+m_{\pi}/M_{\rm d})}=1.0691.
\end {equation}
Minimal coupling corresponds to 
$\omega \to \omega-eV_C$, i.e.,
$$q^2=\omega^2-m_\pi^2  \to (\omega-eV_C)^2 -m_\pi^2 \simeq  q^2-2e\ \omega V_C,$$
 where the Coulomb field $V_C$ originates
 from the extended deuteron charge distribution averaged over the deuteron
 matter distribution. One then has, for the single scattering term, 
the electromagnetic correction 
 
\begin{equation}
\Delta S=-4
\lambda _1\ m_\pi \ b^+e \ \langle  V^d_C(r) \rangle,
\end{equation}

 where , for point particles and in terms of the relative deuteron coordinate {\bf r},
\begin{eqnarray}
 e \langle  V^d_C(r) \rangle = 
 \alpha \times \langle \int  d{\bf r'}
  \frac {\rho (r')}{(|{\bf r}-{\bf r'}|/2)}\rangle \equiv 
2 \alpha  \langle  1/r \rangle_{ch} & = &
 2 \alpha \times \int  \int d{\bf r} d{\bf r'}
 \rho (r) \rho (r')\times (1/|{\bf r}-{\bf r'}|) \nonumber \\
 & = & 4
 \alpha\times \int ^{\infty }_0 dr
 (u(r)^2+w(r)^2) \times (1/r)  \int ^r_0  dr' (u(r')^2+w(r')^2).
\end{eqnarray}

For both  the
Paris~\cite{LAC81} and Bonn2~\cite{MAC87} deuteron 
wave functions
$\langle 1/r \rangle_{ch} $ = 0.300  fm$^{-1}$ and
$e \langle  V^d_C(r) \rangle$ = 0.86 MeV, then 
$$\Delta S = (12 \pm 2) \times 10^{-4}\ m_{\pi }^{-1}.$$

The alternative evaluation, which gauges the $\pi ^-$n interaction
 with the Coulomb 
field from the static spectator proton, gives
\begin{equation}
\Delta S=-2
\lambda _1\ m_\pi\ (b^+-b^-) \ e\langle  V^p_C(r) \rangle
\end{equation}
with
\begin{equation}
 e
 \langle  V^p_C(r) \rangle = \alpha \langle 1/r \rangle=0.66(1)\ {\rm MeV} ,
\end{equation}
using the average inverse deuteron radius of Paris and Bonn2 models, viz.
$\langle 1/r \rangle _d = 0.456(7)$fm$^{-1}$. One obtains,
$$\Delta S = (6 \pm 1) \times 10^{-4}\ m_{\pi }^{-1}.$$

\section{Practical expressions for the theoretical 
$\pi ^-$d scattering length for separable scattering interactions}
\label {Practical expressions for the theoretical pi
-d scattering length for separable scattering interactions}

We give here full practical expressions for the theoretical $\pi ^-$d
scattering length for separable scattering amplitudes with a  dipole
form factor  $v^2(q)=(1+q^2/c^2)^ {-2}$:

\begin{eqnarray}
a_{\pi ^-{\rm d}}~&=& ~\lambda _1
\ (a_{\pi ^-{\rm p}}+ a_{\pi^-{\rm n}})+ \lambda _2 \left [\left (\frac
{a_{\pi ^-{\rm p}}+
a_{\pi^-{\rm n}}}{2}\right )^2-2\left (\frac {a_{\pi ^-{\rm p}}-
a_{\pi^-{\rm n}}}{2}\right )^2\right ]\langle f(r)/r\rangle ~\nonumber\\
&+&~
a({\rm Fermi})~+~a({\rm dispersion })~+~\delta a,
\end{eqnarray}
 where
\begin {equation}
\lambda _2 ~=~2~\frac {(1+m_{\pi}/M)^2}{(1+m_{\pi}/M_{\rm d})}~=~2.4560;
\end {equation}
 \begin{equation}f(r)~=~1-(1+cr/2)\exp(-cr)
\end{equation}
and with the sum of small correction terms
\begin {equation}
\delta a~=~ \delta a({\rm multiple}) + \delta a({\rm isospin})+\delta
a({\rm non-static})+\delta a(\pi ^-{\rm p},\gamma {\rm n})
+\delta a({\rm double ~p-wave})+\delta a({\rm virtual~
pion})\label{eq:deltaa}.
\end{equation}
Assuming isospin symmetry in all terms but the leading order one, and
emphasizing the accurate experimental knowledge of $a_{\pi ^-{\rm p}}$, we have
\begin{eqnarray}
a_{\pi ^-{\rm d}}~&=& ~ \lambda _1
\ (a_{\pi ^-{\rm p}}+ a_{\pi^-{\rm n}})~+  \lambda _2\ \left [a^{+2}-2\ (a_{\pi
^-{\rm p}}-a^+)^2 \right] \langle f(r)/r\rangle ~\nonumber\\
&+&~ a({\rm Fermi})~+~a({\rm dispersion})~+~\delta a.
\label{eq:apimoinsdpracticall}
\end {eqnarray}
The correction for nucleon motion is, according to Eq. (\ref {eq:Fermimotion}):
\begin {equation}
 a({\rm Fermi})~=~2\left (\frac {m_{\pi }}{M+m_{\pi }}\right )^2
 \lambda _1\ c_0\ \left \langle p^2 
\ v^2 \left(\frac{m_{\pi}{\rm p}}{M+m_{\pi }}\right )\right \rangle ,
\end {equation}
where the form factor correction is negligible and $c_0=0.208(3)\
m_{\pi}^{-3}$ ( p.18 in Ref. \cite {ERI88}). The dispersion correction has
been taken to be $a({\rm dispersion})=-56(14)\times 10^{-4}\ m_{\pi
}^{-1}$~\cite{MIZ77}. The remaining values of the small terms are taken to
be (see text) $\delta a({\rm non-static})=11(6)\times 10^{-4}\ m_{\pi
}^{-1}, ~\delta a({\rm double ~p-wave}) = -3\times 10^{-4}\ m_{\pi }^{-1}$
and $\delta a({\rm virtual ~pion}) = -7(2)\times 10^{-4}\ m_{\pi }^{-1}$. In
addition it is desirable to control the convergence of the multiple
scattering expansion explicitly.  We have evaluated the higher order
multiple scattering corrections from the expression given by Kolybasov and
Kudryatsev for the sum of the multiple scattering series to all orders for
point-like scatterers, in the static approximation and neglecting binding
and recoil corrections \cite {KOL80}.  We have however generalized their
expression to include the effect of separable form factors for each
scattering:

\begin {equation}
\delta a ({\rm multiple})~= ~\left \langle \left [2\  \lambda _1
\ a^+~+ \lambda _2
\ \left [a^{+2}-2\ (a_{\pi ^-{\rm p}}-a^+)^2\right]\ \frac {f(r)}{r} \right] 
\left[(1-C)^{-1}-1\right] \right \rangle , 
\label {eq:deltaamultiple}
\end{equation}
where $C=(1+m_{\pi}/M)^2\ [{a^{+2}-2\ (a_{\pi ^-{\rm p}}-a^+)^2]\
f^2(r)/r^2}$. In order to extract the value of $(a_{\pi ^- {\rm p}}+a_{\pi
^+ {\rm p}})/2$ from the experimental $a_{\pi ^-{\rm d}}$ and $a_{\pi ^-{\rm
p}}$, we now observe that Eq.~(\ref{eq:apimoinsdpracticall}) is quadratic in
$a^+$ except for higher power terms from the small $\delta a({\rm
multiple})$ of Eq. (\ref {eq:deltaamultiple}). To check the self-consistency
with $\delta a$(multiple) it
should be solved iteratively. We have done this
with the experimental values and the resulting $a^+$ is small (about
$10^{-3} \ m_{\pi}^{-1}$).  Equation (\ref{eq:apimoinsdpracticall}) can then
be safely linearized for a fixed value of $\delta a$(multiple) and the
consistency checked by iteration. We have

\begin{eqnarray}
\frac {a_{\pi ^-{\rm p}}+a_{\pi ^-{\rm n}}}{2}~&=&
 \left (2\ \lambda _1
 +4\ \lambda _2\ a_{\pi
^-{\rm p}}^{\exp}\ \langle f(r)/r\rangle \right )^{-1}
  \left [a^{exp}_{\pi ^-{\rm d}}+~2\ \lambda _2\ a_{\pi
^-{\rm p}}^{\exp~2}\ \langle f(r)/r\rangle \right.  \nonumber\\
 &-& \left. a({\rm Fermi})~-~a({\rm dispersion})~-~\delta a  \right ] . 
\end{eqnarray}
Two iterations are sufficient.

\pagebreak

\begin{table}
\caption[h]{Some deduced values for the $\pi $NN coupling constant.
The quoted uncertainty are those quoted
by the authors and usually do not include systematic uncertainties.}
\medskip
\begin{tabular}{cccccc}
&Source                            & Year   &System 
&$g^2_{\pi {\rm NN}}/4\pi $&\\
\hline
& Karlsruhe-Helsinki \cite {KOC80}&   1980 &$\pi$p &14.28(18$^*$)&\\
&Kroll \cite {KRO81}              &1981    &pp     &14.52(40$^*$)&\\
\hline  
&Nijmegen \cite {STO93a}          &  1993  & pp, np&13.58~(5$^*$)&\\
&VPI \cite{ARN94a}                &  1994  & pp, np&13.70~~~~~~&\\
&Nijmegen \cite {STO97}           &1997    &pp, np &13.54~(5$^*$)&\\
&Timmermans \cite {TIM97}         &1997    &$\pi^+$p&13.45(14$^*$)&\\ 
\hline
&VPI \cite {ARN94b}  &1994   &GMO,  $\pi$p&13.75(15$^*$)&\\
&Uppsala \cite{RAH98}&1998&np$\rightarrow $pn&14.52(26)&\\
&Pavan {\it et al.} \cite {PAV99}&1999&$\pi$p&13.73~(9)&\\
& Schr\"oder {\it et al.}, corrected \cite{SCH99,WOR92}&1999&GMO,
 $\pi ^{\pm }$p &13.77(18)&\\
&Present work & 2001 &GMO, $\pi ^{\pm }$p &14.11(20)&\\
\end{tabular}
\label{tab:coupling constants}
   $^*$ \begin{small}Statistical uncertainty only.\end {small} 

\end{table}

\begin{table}
\caption[h]{Corrections to $\langle 1/r\rangle $ and to the $\pi $d
scattering length for different cut-off values and wave functions.
The $\pi $N scattering lengths are from 
Eqs.(\ref{eq:a+ determined}) and (\ref{eq:a- determined}).}
\medskip
\begin{tabular}{cccccccc}
&Model&\multicolumn {2}{c}{Paris\cite{LAC81}}&&\multicolumn
{2}{c}{Bonn2\cite{MAC87}}&\\
&$\langle 1/r\rangle $&\multicolumn {2}{c}{0.449
fm$^{-1}$}&&\multicolumn {2}{c}{0.463 fm$^{-1}$}&\\
\hline
&c & $\delta \langle 1/r\rangle $ &$\delta a_{\pi {\rm d}}$ 
&&$\delta \langle 1/r\rangle $&$\delta a_{\pi {\rm d}}$  &\\
&$[m_{\pi}]$&[$10^{-3}\ $fm$^{-1}$]&[$10^{-4}\ m_{\pi}^{-1}$]
&&[$10^{-3}\ $fm$^{-1}$]&[$10^{-4}\ m_{\pi}^{-1}$]&\\
\hline
&3.0 &$-$50& 28&& $-$60&34  &\\
&3.5 &$-$37& 21&& $-$46&26  &\\
&4.0 &$-$28& 16&& $-$36&20  &\\
&4.5 &$-$21& 12&& $-$29&16  &\\
&5.0 &$-$16& ~9&& $-$23&13  &\\
\end{tabular}
\label{tab:form factor corrections}
\end{table}

\begin{table}
\caption[h]{ Estimates of the contribution $ a$(Fermi) to $a_{\pi ^-{\rm d}}$
from single p-wave scattering as a result of Fermi motion  according to
Eq. (\ref{eq:Fermimotion})  for various deuteron wave functions, different
cut-off values and separated into S- and D-state contributions. The last 
row gives $\langle p^2\rangle $ and the kinetic energy 
$\langle p^2\rangle $/M.}
\medskip
\begin{tabular}{cccccccccc}
&Model&\multicolumn {3}{c}{Paris\cite{LAC81}}&&
\multicolumn {3}{c}{Bonn2\cite{MAC87}}&\\
\hline
&c &S-state&D-state &
Total&
&S-state&D-state &Total& \\
&$[m_{\pi }]$&\multicolumn {3}{c}{[in units of $10^{-4}\ m_{\pi }^{-1}$]}
&&\multicolumn {3}{c}{[in units of $10^{-4}\ m_{\pi }^{-1}$]}&\\
\hline
&3&        39.6&27.9&67.6&&36.7&16.7&53.4  &\\
&4&        39.8&28.1&67.8&&36.8&16.9&53.6  &\\
&5&        39.8&28.1&68.0&&36.8&16.9&53.7  &\\
&$\infty $ &39.9&28.3&68.2&&36.8&17.0&53.9  &\\
\hline
\hline 
&$\langle p^2\rangle ~ [$m$_{\pi }^2]$   &0.533&0.378&0.912&&0.492&0.228&0.720\\
&$\langle p^2\rangle /M$ [MeV]&11.1&7.9&19.0&&10.3&4.7&15.0\\
\end{tabular}
\label{tab:p2}
\end{table}

\begin{table}
\caption[h]{ Typical contributions to
$a_{\pi {\rm d}}$ scattering length in units of 10$^{-4}\ m_{\pi}^{-1}$}
\medskip
\begin{tabular}{clccc}
&Contributions&Present work&B--K \cite {BAR97a}&\\
\hline
&$a_{\pi^- {\rm d}}$(double scattering; static)  
              &$-254~(4)^*$& $-252~~~~$&\\
&Fermi motion  &~~~60~(7)   &~~50~~~~&\\
&dispersion correction 
              &$~-56(14)$ &not included&\\
&isospin violation 
               &~~~~3.5~~   &~~~3.5~~&\\
&$(\pi ^-{\rm p},\gamma {\rm n})$ double scattering
              &$~~-2~~~~~$&not considered&\\
&form factor   &~~17~(9)   &~~29(7)~&\\
&higher orders &~~~4~(1)   &~~~6~~~~&\\
&sp interference  &small  &$~-44~~~~$&\\
&non-static effects 
               &~~11~(6)&   ~~10~~~~&\\
&p-wave double scattering  \cite {BAR97a}  
              &$~~-3~~~~~$ 
                          &$~~-3~~~~$&\\
&virtual pion scattering \cite {BEA98,ROB78} 
              &$~~-7~(2)$&not considered&\\ 
\hline
&total = $a_{\pi ^-{\rm d}}-1.07\times (a_{\pi ^-{\rm p}}+a_{\pi ^-{\rm
n}})$ &        $-227(20)$ &$-198~~~~$&\\ 
\hline 
\hline 
&$a_{\pi {\rm d}}$(experimental)\cite{HAU98}& \multicolumn {1}{c}
              {$-252~(7)$}& &\\
\end{tabular}
\label{tab:contributions}
 $^*$ \begin{small}The uncertainty from the $\pi $N scattering 
lengths would typically contribute 
$\pm 6$ units to this term.\end {small} 
\end{table}

\begin{table}
\begin{center}
\caption[h]{Values of $J^-$ from the literature. Quoted errors 
include both statistical and systematic uncertainties.\\}
\medskip
\begin{tabular}{clcc}
 &Source & $J^-$ mb& \\
\hline
 &Koch 1985 \cite {KOC85}                      &$-1.077(47)$  &  \\
& Workman {\it et al.} 1992; K--H \cite {WOR92}&$-1.056~~~~~$& \\
& Workman {\it et al.} 1992; VPI \cite {WOR92} &$-1.072~~~~~$& \\
 &Arndt {\it et al.} 1995 \cite {ARN95}        &$-1.050~~~~~$& \\
&Gibbs {\it et al.} 1998 \cite {GIB98}         &$-1.051(5) ^*$&\\
&Present work                                  &$-1.083(32)$&\\
\end{tabular}
\end{center}
\label {tab:J-values}
 $^*$ \begin{small}statistical error only.\end {small}

\end{table}

\begin{table}
\begin{center}
\caption {Evaluation of $J^-$ in the $\Delta $ resonance region  and up to
2 GeV/c. Here `Data' refers to 'nuclear' 
experimental cross sections uncorrected for Coulomb penetration 
and `Nuclear SM99' to
the corresponding PWA cross sections;
 $I_-$ and $I_+$ are the corresponding integrals for $\pi ^-$p and $\pi ^+$p,
respectively, with $J^-=I_--I_+$. }
\label{Tab. Delta}
\medskip
\begin{tabular}{clccccc}
&Input&$k$ [GeV/c]    & $I_-$ [mb] &$I_+$ [mb]  &$J^-$ [mb] &  \\
\hline
&Hadronic SM95\cite{ARN95b}&0.00~to~0.16&0.164&0.157&~~0.007&\\
&Hadronic SM99\cite{ARN99} &0.00~to~0.16&0.162&0.152&~~0.011&\\
\hline
&Hadronic SM95\cite{ARN95b}&0.16~to~0.55&1.078&2.763&$-1.685$&\\ 
&Hadronic SM99\cite{ARN99} &0.16~to~0.55&1.071&2.767&$-1.696$&\\
&Nuclear SM99\cite{ARN99}  &0.16~to~0.55&1.090&2.726&$-1.636$&\\
&Data\cite {PDG98}         &0.16~to~0.55&1.101&2.753&$-1.652$&\\
\hline
&Hadronic SM95\cite{ARN95b}&0.55~to~1.20&0.800&0.414&~~0.386&\\
&Hadronic SM99\cite{ARN99} &0.55~to~1.20&0.789&0.411&~~0.378&\\
&Nuclear SM99\cite{ARN99}  &0.55~to~1.20&0.804&0.400&~~0.404&\\
&Data\cite{PDG98}          &0.55~to~1.20&0.816&0.400&~~0.416&\\
\hline
&Hadronic SM95\cite{ARN95b}&1.20~to~2.00&0.450&0.460&$-0.010$&\\
&Hadronic  SM99\cite{ARN99}&1.20~to~2.00&0.451&0.458&$-0.007$&\\
&Nuclear SM99\cite{ARN99}  &1.20~to~2.00&0.458&0.450&$~~0.008$&\\
&Data\cite{PDG98}          &1.20~to~2.00&0.459&0.443&~~0.016&\\
\end{tabular}
\end{center}
\label{tab:J- Delta region}
\end{table}

\begin{table} 
\begin{center} 
\caption { Different contributions to $J^-$
as function of the k range and of the input data. The first number in the
parenthesis is the statistical error, while the second numbers correspond
to the systematic uncertainty. The selected data correspond to the world
data as given by PDG Tables, where we have suppressed all data with 
statistical and systematic errors larger than 1\%. Lines
labeled  'Selected' and 'Data' refer to 'nuclear' cross 
sections. }
\label{Tab. Jmoins} 
\medskip 
\begin{tabular}{cclccccc}
&~i & Input & $k$(GeV/c)  & $I_{-}$(mb) 
  & $I_{+}$(mb)       &$J^{-}$(mb)~=~$I_{-} - I_{+}$&\\
\hline
&~1  &   SM95\cite{ARN95b}    & 0.00~to~0.16   &0.164~~~~~~~~~
&0.157~~~~~~~~~~&~0.007~~~~~~~~~&\\
&~2  &   SM99\cite{ARN99}   &     ''           &0.163~~~~~~~~~
&0.152~~~~~~~~~~&~0.011~~~~~~~~~&\\
\hline
&~3  &Selected\cite{PDG98}&0.16 to 2.00        &2.360~(2)~(3)
&3.596~(6)~(1)&$-1.237~(6)~(4)$&\\
&~4  &Data\cite{PED78,CROSSKOCH} &    ''       &2.377~(3)~(2)
&3.596~(5)~(2)&$-1.219~(6)~(4)$&\\
\hline
&~5  &   SM95 \cite{ARN95b}&0.00~to~2.00       &2.492~~~~~~~~~~
&3.794~~~~~~~~~~~&$-1.302~(6)(20)$&\\
&~6  &SM99\cite{ARN99}&   ''                    &2.474~~~~~~~~~~
&3.788~~~~~~~~~~~&$-1.314~(6)(20)$&\\
\hline
&~7  & Selected\cite{PDG98}&2.00 to 4.03 &0.560~(2)~(3)
&0.496~(1)~(5)&~~0.064~(2)~(7)&\\
&~8 &Data\cite{CAR68,CIT66}&''&0.580~(1)~(5)
& 0.518~(1)~(5)&~~0.063~(1)(10)&\\
\hline
&~9  &Selected\cite{PDG98}&4.03 to 240
&2.672~(4)(10)&2.539~(3)(12)&~~0.133~(5)(22)&\\
&10 &Fit PDG94\cite{PDG94}& ''
&2.645~~~~~~~~~~~&2.489~~~~~~~~~~~&~0.155~~~~~~~~~~&\\
\hline 
&11 & Regge 94\cite{PDG94}&240 to$~\infty$ & $-$&  $-$
                            &~~0.030~~~~~~(5)& \\
&12 & Regge 00\cite{GAU00}&'' & $-$ & $-$ 
                            &~~0.025~~~~~~(4)& \\
&13 & Regge 98\cite{PDG98}& '' & $-$ & $-$ 
                            &~~0.018~~~~~~(3)&\\
\hline
&14 & 6+7+10+11&0  to$~\infty$& $-$ &$-$ &$-1.055(10)(31)$&\\
&15 & 6+7+9+11 & ''&  $-$
   &  $-$ &$-1.087~(9)(31)$&\\
&16 & 6+7+9+13 & ''& $-$
   &  $-$ &$-1.099~(8)(31)$&\\
&17 & 6+7+9+12&''& $-$ &$-$&$ -1.092~(9)(31)$&\\
\end{tabular}
\end{center}
\label{Delta contributions}
\end{table}

\end{document}